\documentclass[12pt]{article}
\usepackage{latexsym,epsfig,amssymb}
\newcommand{\ft}[2]{{\textstyle\frac{#1}{#2}}}

\def\bfone{\relax{\rm 1\kern-.35em 1}}

\newcommand{\be}{\begin{equation}}
\newcommand{\ee}{\end{equation}}
\newcommand{\ben}{\begin{displaymath}}
\newcommand{\een}{\end{displaymath}}
\newcommand{\bea}{\begin{eqnarray}}
\newcommand{\eea}{\end{eqnarray}}
\newcommand{\nn}{\nonumber}

\newcommand{\bean}{\begin{eqnarray*}}
\newcommand{\eean}{\end{eqnarray*}}
\newcommand{\beqs}{\begin{eqnarray}}
\newcommand{\eeqs}{\end{eqnarray}}

\newcommand{\mathon}{\mathversion{bold}}
\newcommand{\mathoff}{\mathversion{normal}}

\makeatletter
\@addtoreset{equation}{section}
\makeatother

\begin{document}

\thispagestyle{empty}

\begin{flushright}\small
\end{flushright}

\bigskip
\bigskip

\mathon
\vskip 10mm
\begin{center}
  {\LARGE {\bf $N=8$\, Supergravity}\\[1.2ex]
   {\bf with Local Scaling Symmetry}}
\end{center}
\mathoff


\vskip 8mm

\begin{center}
{\bf Arnaud Le Diffon, Henning Samtleben\footnote{Institut Universitaire de France}}\\[.4ex]
{\small Universit\'e de Lyon, Laboratoire de Physique, UMR 5672, CNRS et ENS de Lyon,\\
46 all\'ee d'Italie, F-69364 Lyon CEDEX 07, France \\
{\tt arnaud.le\_diffon, henning.samtleben @ens-lyon.fr}}
\vskip 8mm
{\bf Mario Trigiante}\\[.4ex]
{\small Dipartimento di Fisica $\&$ INFN, Sezione di Torino, \\
Politecnico di Torino , C.~so Duca degli Abruzzi, 24, I-10129 Torino, Italy \\
{\tt mario.trigiante@polito.it}}
\vskip 4mm
\end{center}

\vskip2cm
\begin{center} {\bf Abstract } \end{center}
\begin{quotation}\noindent
We construct maximal supergravity in four dimensions with local scaling symmetry as
deformation of the original Cremmer-Julia theory.
The different theories which include the standard gaugings
are parametrized by an embedding tensor carrying $56+912$ parameters.
We determine the form of the possible gauge groups and
 work out the complete set of field equations.
As a result we obtain the most general couplings compatible with $N=8$ supersymmetry in four dimensions. A particular feature of these theories is the absence of an action and
 an additional positive contribution to the effective cosmological constant. Moreover, these gaugings are generically dyonic, i.e.\ involve simultaneously electric and magnetic vector fields.
\end{quotation}

\newpage
\setcounter{page}{1}

\tableofcontents

\bigskip
\bigskip

\newpage


\section{Introduction}


$N=8$ supergravity \cite{Cremmer:1979up} is undoubtedly a highly distinguished
field theory due to its high degree of symmetry and the remarkable structure of
its amplitudes that has emerged in recent work, see e.g.~\cite{ArkaniHamed:2008gz,Bern:2009kd}.
The continuous ${E}_{7(7)}$ symmetry underlying the classical field equations
has important consequences for the structure of the counterterms~\cite{Kallosh:2008rr,Brodel:2009hu,Elvang:2010kc,Bossard:2010dq,Beisert:2010jx}.
The field content of maximal supergravity is the unique $N=8$ supermultiplet
\bea
\begin{tabular}{c|rrrrrrrrr}
helicity&$-2$ & $-\frac32$ & $-1$ & $-\frac12$ & $\phantom{-}0$ & $+\frac12$ &$+1$& $+\frac32$ &$+2$ \\\hline
d.o.f. &1&8&28&56&70&56&28&8&1 \\[1ex]
\end{tabular}
\label{tbl:multiplet}
\quad
\;.
\eea
On the other hand, the mutual couplings of the various fields are not uniquely determined,
as supersymmetry allows for the introduction of particular (non-)abelian charges and
the realization of different (non-)abelian gauge groups.
After the original version of the theory with abelian gauge fields \cite{Cremmer:1979up}
the first maximal gauged supergravity was constructed in~\cite{deWit:1982ig}
with the 28 vector fields gauging a compact $SO(8)$ subgroup of ${E}_{7(7)}$.
Non-compact versions of this theory have been constructed and classified
in~\cite{Hull:1984qz,Cordaro:1998tx}
and later been extended to other
non-semisimple gauge groups in~\cite{Andrianopoli:2002mf,Hull:2002cv}.
A general formalism for describing the gauging of subgroups in terms of an
`embedding tensor' has been established in~\cite{deWit:2002vt,deWit:2007mt}.
This constant tensor describes the embedding of the gauge group into the global ${E}_{7(7)}$ symmetry
of the ungauged theory, and parametrizes all the couplings of the gauged theory.

The aim of this paper is the construction of all possible gaugings
(and thus all possible couplings) of $N=8$ supergravity, which in particular
include a gauging of the global scaling symmetry of the theory.
Their gauge groups are embedded in the product $E_{7(7)} \times \mathbb{R}$ of the Cremmer-Julia
group $E_{7(7)}$ with the one-parameter scaling symmetry of the theory that generalizes the Weyl rescaling of general relativity and has been dubbed a `trombone' symmetry of supergravity~\cite{Cremmer:1997xj}.
Supergravity theories that include a gauging of their scaling symmetry have first been constructed
in ten dimensions by a generalized Scherk-Schwarz reduction from eleven dimensions~\cite{Howe:1997qt,Lavrinenko:1997qa}. Lower dimensional examples of such theories include \cite{Bergshoeff:2002nv} and \cite{Kerimo:2003am,Kerimo:2004md}. As a generic feature, these theories are invariant under local rescaling of the fields (including the metric) with appropriate weights upon a compensating gauge transformation on the matter fields. They do not possess an action (since they result from the gauging of an on-shell symmetry) and typically support de Sitter geometries rather than Minkowski or AdS vacua. A systematic account to the construction of these theories has been put forward in~\cite{LeDiffon:2008sh}. Based on the algebraic structure of the duality groups of the ungauged theories, the representation content and the algebraic consistency constraints for the corresponding embedding tensor have been determined for the maximal supergravities. In \cite{Riccioni:2010xx}, these structures have been shown to be naturally embedded in the framework of the very-extended infinite-dimensional Kac-Moody algebra $E_{11}$~\cite{Riccioni:2007au,Bergshoeff:2007qi}.

The general analysis reveals that four-dimensional $N=8$ supergravity admits an embedding tensor transforming in the representation ${\bf 56}+{\bf 912}$ of $E_{7(7)}$, subject to a set of bilinear algebraic consistency constraints. Gaugings defined by an embedding tensor in the irreducible ${\bf 912}$ representation describe gauge groups that entirely reside within $E_{7(7)}$ and have been constructed in \cite{deWit:2002vt,deWit:2007mt}. Additional non-vanishing components in the ${\bf 56}$ representation on the other hand define gaugings that include the trombone generator, i.e.\ theories in which local scaling invariance is part of the gauge group. These are the theories to be constructed in this paper. While the analysis of~\cite{LeDiffon:2008sh} has been purely algebraic and based on the structure of non-abelian deformations of the underlying tensor gauge algebras, it is the aim of this paper to explicitly realize these theories by constructing the full set of supersymmetric field equations. Thereby we derive the most general couplings that are compatible with $N=8$ supersymmetry in four dimensions. In particular, we confirm that the algebraic consistency constraints derived in~\cite{LeDiffon:2008sh} for the embedding tensor are indeed sufficient to ensure supersymmetry of the field equations.

This paper is organized as follows. In section~\ref{sec:gauging}, we analyze the general structure of the gauge groups induced by an embedding tensor in the ${\bf 56}+{\bf 912}$ representation. We explicitly construct the gauge group generators in terms of the embedding tensor and discuss the system of bilinear algebraic consistency constraints that the embedding tensor must satisfy. In case the gauge group includes the trombone generator, this system of constraints drastically reduces upon decomposing the embedding tensor into its $E_{6(6)}$ irreducible components, and we present a number of explicit solutions. We compute the Cartan-Killing metric of the gauge group and show that gaugings involving the local scaling symmetry are generically dyonic, i.e.\ genuinely involve electric and magnetic vector fields.

In section~\ref{sec:coset}, we review the structure of the scalar target space $E_{7(7)}/SU(8)$ and define the $T$ tensor in terms of which the couplings of the gauged theory are expressed.
Subsequently, in section~\ref{sec:susy}, we determine the modified supersymmetry transformations by verifying the closure of their algebra on the bosonic fields of the theory. Based on these results, in section~\ref{sec:eom}, we obtain the modified field equations of the gauged $N=8$ supergravity by starting from an ansatz for the fermionic field equations and calculating their transformation under supersymmetry. This allows to uniquely determine the full set of field equations in lowest order of the fermions. As a particular feature of these theories, we find that gauging of the trombone generator leads to an additional positive contribution to the effective cosmological constant. In section~\ref{subsec:masses}, we determine the conditions for extremality, i.e.\ for solutions of the field equations with constant scalar and gauge fields and give explicit formulas for the mass matrices by linearizing the field equations around these solutions. Finally, in section~\ref{sec:example}, we present a simple example of a theory with local scaling symmetry that has its higher-dimensional origin as a generalized Scherk-Schwarz reduction from five dimensions upon twisting the field with a linear combination of an $E_{6(6)}$ generator and the five-dimensional trombone symmetry. We show that this theory admits a de Sitter solution with constant scalar fields and determine its mass spectrum which seems to indicate that the solution is not stable. We conclude with an outlook on the role and the applications of these theories.


\section{Structure of gauge groups}
\label{sec:gauging}

Before explicitly constructing the full supersymmetric field equations,
in this section we will present and analyze the structure of the possible gauge algebras
that can be realized as local symmetry in maximally supersymmetric supergravity in four dimensions.
Recall, that the global symmetry group
of the ungauged maximally supersymmetric theory is given by~\cite{Cremmer:1979up}
\bea
G &=& E_{7(7)} \times \mathbb{R}
\;,
\label{CJ}
\eea
where the second factor corresponds to the scaling (or trombone) symmetry of the equations of motion,
under which the fields transform as
\bea
\delta g_{\mu\nu} &=& 2\,g_{\mu\nu} \;,\qquad
\delta {\cal A}^M_{\mu} ~=~ {\cal A}^M_{\mu} \;,\qquad
\delta \phi^i ~=~ 0
\;,
\nonumber\\
\delta \psi_\mu &=& \ft12\,\psi_\mu  \;,\qquad
\delta\chi ~=~ -\ft12\,\chi
\;.
\label{scaling}
\eea
Here, the first line refers to the bosonic fields of spin 2, 1, and 0, while the second line gives the transformations of the spin 3/2 gravitons and the spin 1/2 matter fermions, respectively.
The $E_{7(7)}$ factor in (\ref{CJ}) in contrast only acts on vector and scalar fields, with its generators $t_\alpha$
closing into the algebra
\bea
[t_\alpha, t_\beta] &=& f_{\alpha\beta}{}^\gamma\,t_\gamma
\;.
\label{E7algebra}
\eea
The 28 electric vector fields $A_\mu^\Lambda$ combine with their magnetic duals $A_{\mu\,\Lambda}$ into the fundamental 56-dimensional representation $A_\mu^M$ of $E_{7(7)}$ while the 70 scalar fields transform in a non-linear representation parametrizing the coset space $E_{7(7)}/SU(8)$\,.
General gaugings will also require the introduction of two-form tensor fields $B_{\mu\nu\,\alpha}$ transforming in the adjoint 133-dimensional representation of $E_{7(7)}$.

\subsection{Gauge group generators}

In this paper, we will construct the most general supersymmetric theories in which a subgroup of (\ref{CJ}) is gauged. Extending previous work~\cite{deWit:2007mt}, we will consider those theories in which
the gauge groups include the scaling symmetry, i.e.\ the second factor in (\ref{CJ}).

Let us denote by ${\bf t}_{0}$ the generator of the scaling symmetry $\mathbb{R}$, and by
${\cal A}_\mu \equiv \vartheta_M {\cal A}^M_\mu$ the linear combination of vector fields
that will be used to gauge this symmetry upon introduction of covariant derivatives.
As this symmetry also acts in the gravitational sector by scaling the metric, its gauging necessitates
a modification of the spin connection $\omega_{\mu}{}^{ab}$ and the Riemann tensor ${R}_{\mu \nu}{}^{ ab}$ according to~\cite{LeDiffon:2008sh}
\bea
\widehat\omega_{\mu}{}^{ab} &=&\omega_{\mu}{}^{ab}
+2 \,e_\mu{}^{[a}\,\,e^{ b]\, \nu}{\cal A}_{\nu}
\;,
\nonumber\\[1ex]
\widehat{{\cal R}}_{\mu \nu}{}^{ab} &\equiv&
2\,\partial_{[\mu}\,\widehat\omega_{\nu]}{}^{ab} +
2\, \widehat\omega_{[\mu}{}^{ac}\:\widehat\omega_{\nu]c}{}^b
 \nonumber\\[1ex]
&=& {R}_{\mu \nu}{}^{ ab}
- 4 \, e_{[\mu}{}^{[a} \nabla(\omega)_{\nu]}{\cal A}^{b]} + 4 \, e_{[\mu}{}^{[a} {\cal A}_{\nu]} {\cal A}^{b]}
- 2 \, e_{[\mu}{}^{a} e_{\nu]}{}^{b} {\cal A}_{\lambda} {\cal A}^{\lambda}
\;,
\label{modRR}
\eea
which are invariant under the joint transformation
\bea
\delta g_{\mu\nu}  &=& 2 \lambda(x) \,g_{\mu\nu}\;,
\qquad \delta {\cal A}_\mu ~=~ \partial_\mu \lambda(x)
\;.
\eea
Besides, they satisfy the generalized Bianchi identities
\bea
\widehat{{\cal R}}_{[\mu \nu\rho]}{}^{a}
&=&
 {\cal F}_{[\mu\nu}\,e_{\rho]}{}^{a}
\;.
\label{BianchiRR}
\eea

The most general gauging combines this symmetry with some subgroup of the $E_{7(7)}$ factor of (\ref{CJ}).
As shown in \cite{deWit:2002vt,deWit:2007mt} and \cite{LeDiffon:2008sh} for the cases without and
with the trombone factor, respectively, the parametrization of the general gauge group generators
$X_M$ allows for $56+912$ parameters spanning the `embedding tensor' $\vartheta_M$ and $\Theta_M{}^\alpha$, according to
\bea
X_M &\equiv&
 \vartheta_M\,{\bf t}_{0} +
 \left(\Theta_M{}^\alpha + 8 \vartheta_N\,(t^\alpha)_M{}^N\right)\,t_{\alpha}
\;,
\label{generators}
\eea
with covariant derivatives given by $D_\mu \equiv \partial_\mu - {\cal A}^M_\mu X_M$\,.\footnote{
For transparency we have suppressed explicit coupling constants, which can at any stage
be reintroduced by rescaling $\vartheta_M \rightarrow g \vartheta_M$,\,
$\Theta_M{}^\alpha \rightarrow g \Theta_M{}^\alpha$\,.
}
Here, $(t_\alpha)_M{}^N$ are the $E_{7(7)}$ generators (\ref{E7algebra}) in the fundamental
representation,\footnote{
We raise and lower adjoint indices
with the invariant metric $\kappa_{\alpha\beta}\equiv {\rm Tr}\,[t_{\alpha}t_{\beta}]$.
Fundamental indices are raised and lowered with
the symplectic matrix $\Omega^{MN}$ using north-west south-east conventions:
$X^M=\Omega^{MN}X_N$, etc.\,.}
and the matrix $\Theta_M{}^\alpha$ is constrained by the relations
\bea
\Theta_M{}^\alpha (t_\alpha)_N{}^M = 0\;,\qquad
\Theta_M{}^\alpha=-2(t_\beta\, t^\alpha)_M{}^N\,\Theta_N{}^\beta
\;,
\label{linear}
\eea
i.e.\ transforms in the ${\bf 912}$ representation of $E_{7(7)}$.
In absence of $\vartheta_M$, it describes the
gaugings whose gauge group entirely resides within the $E_{7(7)}$.
The relative factors in (\ref{generators}) are chosen such that
the tensor $Z^K{}_{MN} \equiv (X_{(M})_{N)}{}^K$
factors according to
\bea
Z^K{}_{MN} &\equiv& (X_{(M})_{N)}{}^K ~=~
-\ft12 (t_\alpha)_{MN}\left( \Theta^{K\alpha}-16 (t^\alpha)^{KL}\vartheta_L \right)
~\equiv~ (t_\alpha)_{MN}\,Z^{K\alpha}
\;,
\label{defZZ}
\eea
and thus projects onto the ${\bf 133}$ representation in its indices $(MN)$.
This will be a central identity in the construction.
For convenience, we also define the projection
of the gauge group generators onto the $E_{7(7)}$ factor of (\ref{CJ}) as
\bea
\check{X}_M &\equiv&
 \left(\Theta_M{}^\alpha + 8 \vartheta_N\,(t^\alpha)_M{}^N\right)\,t_{\alpha}
\;,
\label{generatorsE7}
\eea

The gauged theory is invariant under the local symmetry
\bea
\delta_\Lambda \vec\phi &=& \Lambda^M X_M \cdot \vec\phi
~\equiv~ \left(\Theta_M{}^\alpha + 8 \vartheta_N\,(t^\alpha)_M{}^N\right) \vec K_\alpha(\phi)
\;,
\nonumber\\
\delta_\Lambda {\cal A}_\mu^M &=& D_\mu \Lambda^M  ~\equiv~ \partial_\mu \Lambda^M
+  {\cal A}^K_\mu (X_{K})_{N}{}^M\,\Lambda^N
\;,
\label{gaugeV1}
\eea
where $\vec K_\alpha(\phi)$ represent the $E_{7(7)}$ Killing vector fields on the scalar target space, and
the gauge group generators are given by evaluating (\ref{generators}) in the appropriate representation, i.e.\
\bea
(X_{K})_{N}{}^M &\equiv&
- \vartheta_K \delta_N^M
+ \left(\Theta_K{}^\alpha + 8 \vartheta_L\,(t^\alpha)_K{}^L\right)\,(t_{\alpha})_N{}^M
\;.
\label{defX}
\eea
Finally, covariant field strengths are defined by
\bea
{\cal H}^M_{\mu\nu} &\equiv&
 2\partial_{[\mu} {\cal A}_{\nu]}^{M}
  + (X_{N})_{P}{}^{M}
  \,{\cal A}_{[\mu}^{N} {\cal A}_{\nu]}^{P}
 + {Z}^{M\alpha}\,B_{\mu\nu\,\alpha}
\;,
\label{defH}
\eea
with a St\"uckelberg-type coupling to the two-forms $B_{\mu\nu\,\alpha}$ and
the (constant) intertwining tensor ${Z}^{M\alpha}$ defined in (\ref{defZZ}).
They transform covariantly under the gauge transformations (\ref{gaugeV1})
provided the two-forms transform as
\bea
\delta_\Lambda B_{\mu\nu\,\alpha} &=&
-2(t_\alpha)_{MN}\,\left(\Lambda^M {\cal H}^N_{\mu\nu}
- {\cal A}_{[\mu}^M\, \delta {\cal A}_{\nu]}^N\right)
\;.
\label{gaugeV2}
\eea
Moreover, the covariant field strengths (\ref{defH}) are invariant under the tensor
gauge transformations\footnote{
W.r.t.\ reference~\cite{deWit:2007mt} we have rescaled the two-form fields and associated tensors as
$B_{\mu\nu\alpha}\rightarrow-B_{\mu\nu\alpha}$,
$\Xi_{\mu\alpha}\rightarrow-\Xi_{\mu\alpha}$,
$Z^{M\alpha} \rightarrow -Z^{M\alpha}$.}
\bea
\delta_\Xi B_{\mu\nu\,\alpha} &=&
2{\cal D}_{[\mu} \Xi_{\nu]\,\alpha} + 2(t_\alpha)_{MN}\,{\cal A}_{[\mu}^M\, \delta {\cal A}_{\nu]}^N
\;,\nonumber\\
\delta_\Xi {\cal A}_\mu^M &=& - Z^{M\alpha}\, \Xi_{\mu\,\alpha}
\;.
\label{gaugeT}
\eea
The covariant field strengths (\ref{defH}) satisfy the
generalized Bianchi identities
\bea
{\cal D}^{\vphantom{M}}_{[\mu}{\cal H}^M_{\nu\rho]} &=&
\ft13 {Z}^{M\alpha}\,{\cal H}_{\mu\nu\rho\,\alpha}
\label{Bianchi}
\;,
\eea
with the covariant non-abelian field strength ${\cal H}_{\mu\nu\rho\,\alpha}$
of the two-form tensor fields, given by:
\begin{eqnarray}
{\cal H}_{\mu\nu\rho\,\alpha}&=&3 D_{[\mu}B_{\nu\rho]\alpha}+6\,t_{\alpha PQ}\,A^P_{[\mu}\left(\partial_\nu A^Q_{\rho]}+\frac{1}{3}\,X_{RS}{}^Q\,A^R_\nu A^S_{\rho]}\right)\,,
\end{eqnarray}
where
\begin{eqnarray}
D_{[\mu}B_{\nu\rho]\alpha}&=&\partial_{[\mu}B_{\nu\rho]\alpha}+2\,t_{\alpha PQ} Z^{Q\beta}A^P_{[\mu}B_{\nu\rho]\beta}\,.
\end{eqnarray}

\subsection{Consistency constraints}
\label{concon}

The previous construction leads to a consistent (closed) gauge algebra, if the irreducible components
$\vartheta_M$, $\Theta_M{}^\alpha$ satisfy the following system of quadratic constraints~\cite{LeDiffon:2008sh}
\bea
\vartheta_{M}\, \Theta^M{}^\alpha
&\stackrel!{\equiv}&16\, (t^\alpha)^{MN} \,\vartheta_{ M}\, \vartheta_{ N}
\;,
\label{Q1}\\
 (t_{\gamma})_{[M}{}^{ P}\, \Theta_{N]}{}^\gamma\,\vartheta_{ P}
&\stackrel!{\equiv}&0
\;,
\label{Q2}\\
\Theta_M{}^\alpha\,\Theta^M{}^\beta  &\stackrel!{\equiv}&
8 \,\vartheta_M\,\Theta_N{}^{[\alpha}\,t^{\beta]}{}^{MN}
-4\,f^{\alpha\beta}{}_{\gamma}\,\vartheta_M\,\Theta^M{}^\gamma
\;,
\label{Q3}
\eea
transforming in the ${\bf 133}$, the ${\bf 1539}$ and the ${\bf 133}+{\bf 8645}$,
of $E_{7(7)}$, respectively.
For $\vartheta_M=0$ they consistently reduce to the quadratic conditions of~\cite{deWit:2007mt}.
As we will show in the following, any solution to the constraints
(\ref{Q1})--(\ref{Q3}), will define a viable maximally supersymmetric gauged supergravity.

It is straightforward to show that (\ref{Q1})--(\ref{Q3}) imply several direct consequences
for the gauge group generators, such as the closure of the gauge algebra according to
\bea
{}[X_M,X_N] &=& -X_{MN}{}^K\,X_K \;,
\label{XXX}
\eea
and orthogonality between gauge group generators and the
intertwining tensor $Z$
\bea
X_{MN}{}^K\,Z^{M\alpha}  ~=~ 0 ~=~
\vartheta_M\,Z^{M\alpha}
\;.
\label{QQZ}
\eea
The reason for the fact that the gauge transformations consistently
close into an algebra when properly extended to the two-form tensor fields
even in presence of the gauging of the scaling symmetry
is the underlying structure of a hierarchy of non-abelian tensor
gauge transformations~\cite{deWit:2005hv,deWit:2008ta} which is not based on the
existence of an action. The relative factors in (\ref{generators}) and the identity (\ref{defZZ})
are central in this construction.
What we will show explicitly in this paper is that the non-abelian deformations
defined in the previous section are precisely the ones that are moreover compatible with maximal
supersymmetry of the field equations.

\subsection{Solution to the quadratic constraints}
\label{sttqc}

In general, it is a hard task to construct solutions to the quadratic constraints of gauged supergravity.
However, it turns out that in presence of the trombone (i.e.\ non-vanishing $\vartheta_M$), the system
(\ref{Q1})--(\ref{Q3}) can be reduced to a much simpler one in terms of a reduced number of
components. The strategy for solving the quadratic constraints follows the case of pure trombone
gaugings~\cite{LeDiffon:2008sh} by decomposing all objects with respect to the $E_{6(6)}\times SO(1,1)$
subgroup of $E_{7(7)}$.
Explicitly, this means that the adjoint representation branches as
\bea
t_\alpha &\rightarrow& (t_{\rm o},\; t_a,\; t_m,\; t^m)
\;,\nonumber\\
{\rm according~to}\;\;
{\bf 133} &\rightarrow& {\bf 1}^{0}+{\bf 78}^{0}+{\bf 27}^{-2}+\bar{\bf 27}^{+2}
\;,
\eea
while the fundamental representation breaks into
\bea
\vartheta_M&\rightarrow~&
(\vartheta_\bullet,\;
\vartheta_m,\;
\vartheta^m,\;
\vartheta^\bullet)
\;,
\nonumber\\
{\rm according~to}\;\;
{\bf 56} &\rightarrow& {\bf 1}^{+3}+{\bf 27}^{+1}+\bar{\bf 27}^{-1}+{\bf 1}^{-3}
\;,
\eea
and the embedding tensor $\Theta_M{}^\alpha$ decomposes into
\bea
\Theta_M{}^\alpha &\rightarrow~&
\left( \xi_+^{a},\;
\xi_m,\;
\xi^{mn},\;
\xi_{mn},\;
\xi^m,\;
 \xi_-^{a}
\right)
\;,
\nonumber\\
{\rm according~to}\;\;
{\bf 912} &\rightarrow&
{\bf 78}^{+3}+{\bf 27}^{+1}+\overline{\bf 351}^{+1}
+{\bf 351}^{-1}+\bar{\bf 27}^{-1}+{\bf 78}^{-3}
\;,
\eea
with its explicit $(56\times 133)$ matrix form given in (\ref{ThetaBreak}) in the appendix.
We use indices $a, b, \dots = 1, \dots, 78$
and $m, n, \dots = 1, \dots, 27$ to label the adjoint and the fundamental representation of $E_{6(6)}$,
respectively.

In appendix~\ref{app:solve} we derive an important result: for non-vanishing $\vartheta_M$ and
up to $E_{7(7)}$ rotations, the general
solution to the system (\ref{Q1})--(\ref{Q3}) is parametrized by a real constant $\kappa$, an $E_{6(6)}$ matrix
$\Xi_m{}^n \equiv \Xi^a (t_a)_m^n$ and two real tensors $\zeta^m$, $\zeta^{[mn]}$, as follows
\bea
(\vartheta_\bullet,\vartheta_n,\vartheta^n,\vartheta^\bullet)&=&
 (\kappa,0,\kappa\zeta^n,0)
\;,
\nonumber\\
\left( \xi_+^{a},\;
\xi_m,\;
\xi^{mn},\;
\xi_{mn},\;
\xi^m,\;
 \xi_-^{a}
\right)&=&
\left(\Xi^a, 0,\zeta^{mn}, \Xi_{[m}{}^k d_{n]kl} \zeta^l ,-\ft43 \kappa \zeta^m, 0\right)
\;,
\label{solutionemb}
\eea
where $d_{kmn}$ denotes the totally symmetric $E_{6(6)}$ invariant tensor.
The tensors $\zeta^m$, $\zeta^{[mn]}$ must be real eigenvectors under the action of $\Xi$ according to
\bea
\delta_\Xi \zeta^m &\equiv& - \Xi_n{}^m \zeta^n ~\stackrel!{\equiv}~ \ft43\kappa \zeta^m \;,\nonumber\\
\delta_\Xi \zeta^{mn} &\equiv& 2\Xi_k{}^{[m} \zeta^{n]k} ~\stackrel!{\equiv}~ \ft23\kappa \zeta^{mn}
\;,
\label{remcon}
\eea
which furthermore must satisfy the following set of polynomial constraints:
\bea
 \zeta^k\zeta^l d_{mkl} &\stackrel!{\equiv}& 0
\;,\label{cs1}\\
\zeta^{k} \zeta^{mn} d_{kml} &\stackrel!{\equiv}& 0\;,
\label{cs2}\\
\zeta^{[k} \zeta^{mn]} &\stackrel!{\equiv}& 0 \;,
\label{cs3}\\
\left(t_a \cdot (\Xi+ \ft43 \kappa I) \cdot (\Xi-\ft23 \kappa I)\right){}\!_n{}^m \,\zeta^n &\stackrel!{\equiv}&
-\ft12 \zeta^{mk} \zeta^{ln} d_{klp} (t_a)_n{}^p
\;.
\label{cs4}
\eea
As we show in appendix~\ref{app:solve}, this system of equations
is equivalent to the original system of constraints~(\ref{Q1})--(\ref{Q3}).
In contrast to the original system, solutions to (\ref{remcon})--(\ref{cs4}) may easily be constructed.

A simple solution to the system (\ref{remcon})--(\ref{cs4}) is given by setting $\zeta^m=0=\zeta^{mn}$\,.
This leaves a non-trivial embedding tensor (\ref{solutionemb}) parametrized by $\kappa$
and an $E_{6(6)}$
generator $\Xi$.
This solution satisfies a stronger version of the quadratic constraints:
left and right hand sides of equations (\ref{Q1})--(\ref{Q3}) vanish separately.
In the limit $\kappa\rightarrow0$ in which $\vartheta_M$ vanishes,
this solution corresponds to the
known gaugings induced by a Scherk-Schwarz reduction~\cite{Scherk:1979zr}
from five dimensions parametrized by the choice of
an $E_{6(6)}$ generator~\cite{Andrianopoli:2002mf}.
For non-vanishing $\kappa$, the higher-dimensional origin of these
theories is a generalized Scherk-Schwarz reduction
from five dimensions in which the fields are twisted by a linear combination of the $E_{6(6)}$
generator $\Xi$ and
the five-dimensional trombone symmetry. The form of the generators (\ref{generators})
shows that even for vanishing $\Xi=0$,
switching on $\kappa$ corresponds to gauging a linear combination of the four-dimensional
trombone generator ${\bf t}_0$ and a subset of $E_{7(7)}$ generators.
More complicated solutions of the constraints involve
non-vanishing zero-modes $\zeta^m,\,\zeta^{mn}$.
While we defer the complete solution of the constraint
system (\ref{remcon})--(\ref{cs4})  to a separate publication, a typical example of such a solution will be
discussed in Section \ref{awoe}.

\subsection{Invariants of the trombone}
%
We can classify the inequivalent gaugings according to the ${\rm
E}_{7(7)}$-invariants constructed out of the embedding tensor. In
particular, the quadratic constraints can be regarded as conditions
on the ${\rm E}_{7(7)}$-orbits of the embedding tensor. In terms of
$\vartheta_M$ and $\Theta_M{}^\alpha$, several ${\rm
E}_{7(7)}$-invariants can be constructed of which the simplest is
the quartic invariant $I_4(\vartheta)$ depending only on the
trombone component $\vartheta_M$ according to
\begin{eqnarray}
I_4(\vartheta)&\equiv& -2\,(t_{\alpha})^{MN}(t^\alpha){}^{PQ}\:
\vartheta_M\,\vartheta_N\,\vartheta_P\,\vartheta_Q\nonumber\\[1ex]
&=&
-(\vartheta_\bullet\,\vartheta^\bullet+\vartheta_m\,\vartheta^m)^2+10
d_{mnp}\,d^{mrs}\vartheta_r\vartheta_s\,\vartheta^n\,\vartheta^p
\nonumber\\
&&{} -\ft{20}{3}\,
\vartheta^\bullet\,d^{mnp}\,\vartheta_m\vartheta_n\vartheta_p+\ft{2}{3}\,
\,\vartheta_\bullet\,d_{mnp}\,\vartheta^m\vartheta^n\vartheta^p\,.
\label{I4}
\end{eqnarray}
The different orbits of the 56-dimensional fundamental
representation of $E_{7(7)}$ are characterized via this invariant
as~\cite{Ferrara:1997uz}:
\begin{itemize}

\item[$(i)$]
{$I_4(\vartheta)>0$: the orbit is $\frac{{\rm E}_{7(7)}}{{\rm
E}_{6(2)}}$\,;}

\item[$(ii)$]
{$I_4(\vartheta)<0$: the orbit is $\frac{{\rm E}_{7(7)}}{{\rm
E}_{6(6)}}$\,;}

\item[$(iii)$]
{$I_4(\vartheta)=0,\,\frac{\partial I_4(\vartheta)}{\partial
\vartheta_M}\neq 0$:
 the orbit is $\frac{{\rm E}_{7(7)}}{{\rm F}_{4(4)}\ltimes T_{26}}$\,;}

 \item[$(iv)$]
 {$I_4(\vartheta)=0,\,\frac{\partial I_4(\vartheta)}{\partial \vartheta_M}= 0,\,
 t_{\alpha\,MN}\vartheta^M\,\vartheta^N\neq0$:
 the orbit is $\frac{{\rm E}_{7(7)}}{{\rm SO}(6,5)\ltimes (T_{32}\times T_1)}$\,;}

 \item[$(v)$]
 {$ t_{\alpha\,MN}\vartheta^M\,\vartheta^N=0$:
 the orbit is $\frac{{\rm E}_{7(7)}}{{\rm E}_{6(6)}\ltimes T_{27}}$\,.}

\end{itemize}
Inserting the solution (\ref{solutionemb}) obtained in the previous section
into (\ref{I4}), we find
\begin{eqnarray}
I_4(\vartheta)&=&\frac{2}{3}
\,\kappa^4\,d_{mnp}\,\zeta^m\zeta^n\zeta^p\,.
\end{eqnarray}
From (\ref{cs1}) it follows that
$I_4(\vartheta)=0=\,\frac{\partial I_4(\vartheta)}{\partial
\vartheta_M}$. Since $t_{\alpha\,MN}\vartheta^M\vartheta^N$ has a
non-vanishing component $(t^m)^\bullet{}_n\,\vartheta_\bullet
\vartheta^n\propto \kappa^2\,\zeta^m$, we conclude that
$\vartheta_M$ belongs to the orbit~$(iv)$ of the above classification if $\zeta^k$ is
non-vanishing, and otherwise to the orbit~$(v)$.\par In both cases
the gauge group $G_g$ will then be a subgroup of the stability
group of the corresponding orbit inside $\mathbb{R}^+\times{\rm
E}_{7(7)}$.  In case $(iv)$, for instance, we should have
\begin{eqnarray}
G_g&\subset&\left[\mathbb{R}^+\times{\rm SO}(6,5)\right]\ltimes
(T_{32}\times T_1)\,,
\end{eqnarray}
where $\mathbb{R}^+$ is a suitable combination of the trombone
symmetry and the ${\rm O}(1,1)^7$ symmetry inside ${\rm E}_{7(7)}$.

\subsection{An explicit example}\label{awoe}
%
Here we present an example of a solution of the constraints
 (\ref{remcon})--(\ref{cs4}).
The quadratic  condition (\ref{cs1}) on $\zeta^m$:
\begin{eqnarray}
d_{mnp}\,\zeta^n\,\zeta^p&=&0\,,
\end{eqnarray}
can be viewed as a kind of ``$ {\rm E}_{6(6)}$-pure spinor''
constraint. It defines an orbit of the $\overline{{\bf 27}}$ with
stability group ${\rm SO}(5,5)\ltimes T_{16}$ (see
\cite{Ferrara:1997uz}). This means that there exists an ${\rm
SO}(5,5)\subset {\rm E}_{6(6)} $ with respect to which $\zeta^m$ is
a singlet. If we decompose the adjoint and the fundamental
representations of ${\rm E}_{6(6)}$ with respect to ${\rm O}(1,1)\times
{\rm SO}(5,5)\subset {\rm E}_{6(6)}$ we find:
\begin{eqnarray}
{\bf 78}&\rightarrow & {\bf 1}^0+{\bf 45}^0+{\bf 16}_c^{+3}+{\bf 16}_s^{-3}\,,\nonumber\\
\overline{{\bf 27}}&\rightarrow & {\bf 1}^{-4}+{\bf 10}^{+2}+{\bf
16}_c^{-1}\,.\label{o11o55branch}
\end{eqnarray}
The stabilizer of $\zeta^m$ is thus generated by the ${\bf
45}^0+{\bf 16}_s^{-3}$, while $\zeta^m$ corresponds to the ${\bf
1}^{-4}$.  We denote by ${\bf h}$ the ${\rm O}(1,1)$ generator,
such that $\delta_{\bf h}\, \zeta^m=-4\,\zeta^m$.
Eqs.~(\ref{remcon}), on the other hand, imply that $\delta_\Xi\,
\zeta^m=-\Xi_n{}^m\,\zeta^n=\frac{4}{3}\,\kappa\,\zeta^m$. Since
$\zeta^m$ is a simultaneous eigenvector of both $\Xi$ and ${\bf h}$,
we must have $\delta_{[\Xi,\,{\bf h}]}\, \zeta^m=0$, namely that
$\Xi$ cannot have a component along the ${\bf 16}_c^{+3}$:
\begin{eqnarray}
\Xi &\in & {\bf 1}^0+{\bf 45}^0+{\bf 16}_s^{-3}\,.
\end{eqnarray}
We conclude that $\Xi$ consists of a component proportional
to ${\bf h}$ plus an element $\Xi_0$ in the algebra of the little
group of $\zeta^m$:
\begin{eqnarray}
\Xi &=& -\frac{1}{3}\,k\,{\bf h}+\Xi_0\;,\qquad
\delta_{\Xi_0} \zeta^m=0\,.\label{xisemisimple}
\end{eqnarray}
Let us consider the case in which $\Xi_0$ is a semisimple element of
$\mathfrak{so}(5,5)$ and thus can be taken as an element of its
Cartan subalgebra.  One can show that in this case, taking $\zeta^{mn}=
\zeta^{[m} \eta^{n]}$, with $\eta^m$ in the ${\bf 16}_c^{-1}$, all the
constraints are satisfied. In particular the two sides of Eq.~(\ref{cs4}) are separately zero.
As we shall show in Appendix~\ref{etre}, this equation in particular implies that $\Xi_0$ should
commute with an ${\rm SO}(4,4)$ subgroup of ${\rm SO}(5,5)$.
 The resulting gauge algebra $\mathfrak{g}_g$ is
$21$-dimensional and of the form:
\begin{eqnarray}
\mathfrak{g}_g&=&\mathfrak{o}(1,1)\oplus\mathfrak{so}(2,1)\oplus\mathfrak{l}^{(2\kappa)}\oplus\mathfrak{l}^{(4\kappa)}
\;,\nonumber\\[1ex]
&&{\rm dim}(\mathfrak{l}^{(2\kappa)})=16\;,\quad
{\rm dim}(\mathfrak{l}^{(4\kappa)})=1\,,\label{ga}
\end{eqnarray}
the gradings referring to the ${\rm O}(1,1)$-generator. We can
understand the embedding of the gauge group into the stability group
$\left[\mathbb{R}^+\times{\rm SO}(6,5)\right]\ltimes
(T_{32}\times T_1)$ of the
$\vartheta_M$-orbit by decomposing ${\rm SO}(6,5)$ with respect to
the ${\rm SO}(2,1)\times {\rm SO}(4,4)$. Then the generators of
$\mathbb{R}^+\times {\rm SO}(2,1)$ provide the zero-grading part of
the gauge algebra (\ref{ga}). The gauge generators can be
written in a manifestly ${\rm SO}(2,1)\times {\rm
SO}(4,4)$-covariant way. Let $A,B=1,2$ denote the ${\rm
SO}(2,1)$-doublet indices while  $I,J=1,\dots, 8$ label the ${\bf
8}_s$ of ${\rm SO}(4,4)$. Then let $T_x$, $x=0,1,2,3$, be the
$\mathfrak{o}(1,1)+\mathfrak{so}(2,1)$ generators,
$\mathfrak{l}^{(2\kappa)}={\rm Span}(T_{AI})$ and
$\mathfrak{l}^{(4\kappa)}={\rm Span}(T)$. The relevant commutation
relations between the gauge generators are:
\begin{eqnarray}
[T_x,\,T_{AI}]&=&-(T_x)_A{}^B\,T_{BI}\;,
\qquad
[T_{AI},\,T_{BJ}]=\epsilon_{AB}\,C_{IJ}\,T\;,
\label{gaugecomm}
\end{eqnarray}
where $C_{IJ}$ is the symmetric invariant matrix in the product
${\bf 8}_s\times {\bf 8}_s$.
 In other words, with respect to ${\rm SO}(1,1)\times{\rm
SO}(2,1)\times {\rm SO}(4,4)$ the generators $\{T_x\}$ are in the
${\bf (3,1)}^0$, $\{T_{AI}\}$ in the ${\bf (2,8_s)}^{2\kappa}$ while
$\{T\}$ is in the ${\bf (1,1)}^{4\kappa}$. In terms of the ${\rm
E}_{7(7)}$-branching with respect to the ${\rm E}_{6(6)}$-subgroup,
the $T_{AI}$ consists of $8$ generators from the $\overline{{\bf
27}}$ and $8$ from the ${\bf 78}$, while $T$ originates from the
$\overline{{\bf 27}}$.
 This structure does not change either in the limit
 $\zeta^{mn}\rightarrow 0$, or in the limit
$\zeta^m\rightarrow 0$. In the latter case the $\mathfrak{gl}(2)$
algebra of ${\rm SO}(1,1)\times{\rm SO}(2,1)$ contracts to a
non-semisimple  algebra of the form $\mathfrak{o}(1,1)+H_3$, where
$H_3$ is a three-dimensional Heisenberg algebra spanned by nilpotent
generators. Only if both the zero-modes vanish
($\zeta^{mn}\rightarrow 0$, $\zeta^m\rightarrow 0$) the $T_{AI}$
generators which do not vanish become abelian, the last commutator in
(\ref{gaugecomm}) becomes trivial and we retrieve the first example
discussed in section~\ref{sec:example}.

%
\subsection{Cartan-Killing metric of the gauge group}
%

In the previous sections we have been discussing the general
solution to the quadratic constraints and worked out the
corresponding gauge groups in certain examples. With the general solution given
in section \ref{sttqc}, the gauge group generators may be reconstructed from (\ref{defX}),
putting together (\ref{thbreak}), (\ref{ThetaBreak}) and (\ref{solutionemb}). The explicit form of
the generators $\{X_M\} = \{X_\bullet, X_m, X^m, X^\bullet\}$ in terms of the parameters
$\kappa$, $\Xi^a$, $\zeta^m$, and
$\zeta^{mn}$ is given in (\ref{explicitgenerators}) in the appendix.
Via (\ref{XXX}) these generators also encode the structure constants
of the gauge algebra. We can compute the Cartan-Killing
metric of the gauge group as $g_{MN}\equiv{\rm Tr}(X_M\,X_N)$. Its
non-vanishing components are \bea g_{\bullet \bullet} &=&
64\kappa^2+ 2\,\Xi^a \Xi_{a}
\;,\nonumber\\
g_{\bullet}{}^m &=& (96\kappa^2- 2\,\Xi^a \Xi_{a}) \,\zeta^m
\;,\nonumber\\
g_{\bullet}{}_m &=& -6\,\Xi_l{}^n d_{nmk} \zeta^{kl}
\;,\nonumber\\
g^{mn} &=& (64\kappa^2+ 2\,\Xi^a \Xi_{a}) \,\zeta^m\zeta^n
\;,\nonumber\\
g_{mn} &=& -6\, \zeta^{kl}\zeta^{pq} d_{mkp} d_{nlq}
\nonumber\\
&&{} +\ft23\,(80\kappa^2-3\,\Xi^a \Xi_{a})\,d_{mnk}\,\zeta^k +
24\,d_{kl(m}\,(\Xi^2)_{n)}{}^k \zeta^l \;.
\eea
For $\zeta^m=0=\zeta^{mn}$ this shows that the semisimple part
of the gauge algebra is one-dimensional in accordance with its origin as
a Scherk-Schwarz reduction from five dimensions.
If $\zeta^{mn}=0$ but $\zeta^m$ is non-vanishing, a contraction of equation (\ref{cs4})
implies that $\Xi^a \Xi_{a}=32\kappa^2$, and the Cartan-Killing form accordingly reduces to
\bea
g_{\bullet \bullet} &=& 128\kappa^2
\;,\nonumber\\
g_{\bullet}{}^m &=& 32\kappa^2 \,\zeta^m
\;,\nonumber\\
g^{mn} &=& 128 \kappa^2\,\zeta^m\zeta^n
\;,\nonumber\\
g_{mn} &=& -\ft{32}3\kappa^2 \,d_{mnk}\,\zeta^k
+ 24\,d_{kl(m}\,(\Xi^2)_{n)}{}^k \zeta^l
\;.
\eea

%
\subsection{Electric/magnetic gaugings}
%

So far, we have discussed the structure of the gauge algebra by studying
deformations that involve vector fields from the entire 56-dimensional
fundamental representation of $E_{7(7)}$.
It is well known~\cite{Cremmer:1979up},
that only half of these vector fields are dynamical electric vector fields
while the other half is given by their magnetic duals. Accordingly, only the former half
appears in the action of the ungauged theory.
Nevertheless, the connections of a general gauging may contain magnetic
vector fields that are related by their first order duality equations to the electric fields of the theory.
In~\cite{deWit:2005ub} it has been shown how to elevate this construction to the level
of an action by introducing additional auxiliary two-form tensor fields
(which in turn are the magnetic duals to the scalar fields of the theory).
The magnetic vector fields then do not possess a standard kinetic term but rather couple via
a topological $BF$ term to the two-form tensor fields.
On the other hand, all standard gaugings of the theory~\cite{deWit:2007mt}
satisfy a symplectic locality condition that ensures the existence of a symplectic frame
in which all the vector fields involved in the gauging live in the electric sector.
In this sense even in presence of magnetic charges these theories
remain electric gaugings in disguise
which is in accordance with general results on the gauging of
electric/magnetic duality~\cite{Bunster:2010wv,Deser:2010it}.
We shall see that this is no longer the case for the gaugings considered in this paper,
related to the fact that these theories do no longer admit an action.

For the solution of the consistency constraints of the embedding tensor
discussed at the end of section~\ref{sttqc}, the
left and right hand sides of equations (\ref{Q1})--(\ref{Q3}) vanish separately.
The gauge group generators thus satisfy the symplectic locality condition
$\Omega^{MN} X_M X_N  = 0$. I.e.\
as for the standard gaugings we can choose a symplectic frame
$\{X^M\} \rightarrow \{X^\Lambda, X_\Lambda\}$ such that all $X^\Lambda$ are identically zero.
Indeed, in this case the explicit form of the generators (\ref{explicitgenerators})
shows that $X_\bullet=0=X^k$\,.
Accordingly, the gauging only involves electric vector fields $\{{\cal A}_\mu^\Lambda\} = \{{\cal A}_{\mu\,\bullet}\,, {\cal A}_\mu^k\}$.
On the other hand, in the generic case the components
$\zeta^m$, $\zeta^{mn}$ in the embedding tensor are non-vanishing,
such as in the example worked out in section~\ref{awoe}. Then, equation~(\ref{Q3})
implies that $\Omega^{MN} X_M X_N  \not=0$, i.e.\ there is no symplectic frame
in which the gauging involves only electric vector fields.
We conclude that the general gaugings including the trombone generator
are necessarily and genuinely dyonic!

%
\mathon
\section{Scalar coset space and the $T$-tensor}
\mathoff
\label{sec:coset}


In this section, we discuss the structure of the scalar sector of the theory,
discuss its interplay with the gauging defined in the previous section,
and define the relevant scalar field dependent tensors ($T$-tensors) that enter
in the field equations of the gauged supergravity.
%
%
\mathon
\subsection{Coset space $E_{7(7)}/SU(8)$}
\mathoff
%

The scalar fields of $N=8$ supergravity
can be parametrized in terms of the 56-dimensional complex vectors
${\cal V}_M{}^{ij}= ({\cal V}_\Lambda{}^{ij}, {\cal V}^{\Sigma \,ij})$ and their
complex conjugate ${\cal V}_{M\,ij}=({\cal V}_{\Lambda\,ij},{\cal V}^\Sigma{}_{ij})$,
which together constitute a $56\times 56$ matrix ${\cal V}$,
\begin{equation}
  \label{VV}
  {\cal V}_M{}^{\underline N} =\Big({\cal V}_M{}^{ij}, {\cal V}_M{}_{kl} \Big) =
  \pmatrix{{\cal V}_\Lambda{}^{ij}&{\cal V}_{\Lambda\,kl}\cr
\noalign{\vskip 4mm}
       {\cal V}^{\Sigma\,ij} & {\cal V}^\Sigma{}_{kl}\cr}\,.
\end{equation}
Indices $M, N, \dots = 1, \dots, 56$, label the fundamental representation of $E_{7(7)}$,
indices $i, j, \dots = 1, \dots, 8$ denote the fundamental complex ${\bf 8}$ of $SU(8)$.\footnote{
Earlier, in section~\ref{sttqc} we have used indices $m, n, \dots$ in a different context labeling
the 27 dimensional fundamental representation of $E_{6(6)}$. We hope that this does not
lead to extra confusion.}

The underlined indices $\underline{M}, \underline{N}, \dots = 1, \dots, 56$, are a collective label for
the ${\bf 28}$ + $\bar{\bf 28}$ of $SU(8)$\,.
The matrix ${\cal V}_M{}^{\underline N}$
transforms under rigid $E_{7(7)}$ from the left and under
local ${SU}(8)$ from the right. Strictly speaking, it does not constitute an
element of $E_{7(7)}$, but it is equal to a constant matrix (to account for
the different bases adopted on both sides) times a space-time
dependent element of $E_{7(7)}$.
We refer to~\cite{deWit:1982ig,deWit:2007mt} for further details.
In particular, the scalar matrix satisfies the properties
\begin{eqnarray}
  \label{eq:VV-orthogonal}
  {\cal V}_M{}^{ij} \,{\cal V}_{N\,ij} - {\cal V}_{M\,ij}\, {\cal V}_N{}^{ij}  &=&
  \mathrm{i}\,\Omega_{MN}\,, \nonumber\\
  \Omega^{MN} \,{\cal V}_M{}^{ij} \,{\cal V}_{N\,kl} &=&
  \mathrm{i}\,\delta^{ij}{}_{kl}\,, \nonumber\\
  \Omega^{MN} \,{\cal V}_M{}^{ij} \, {\cal V}_N{}^{kl} &=& 0\,,
\end{eqnarray}
reflecting the fact that $E_{7(7)}$ is embedded into $Sp(56)$.
The covariant scalar currents ${\cal Q}_{\mu}{}_{i}{}^{j}$ and $\mathcal{P}_\mu{}^{ijkl}$ are defined by
\bea
\partial_\mu{\cal V}_M{}^{ij}
  - A_\mu^P X_{PM}{}^N \,{\cal V}_N{}^{ij}
&\equiv& - \mathcal{Q}_{\mu \,k}{}^{[i} \,{\cal V}_M{}^{j]k}
+ \mathcal{P}_\mu{}^{ijkl} \,{\cal V}_{M kl}
\;,
\label{defQP}
\eea
with gauge group generators from (\ref{defX}), and satisfy
\bea
{\cal Q}_{\mu}{}^{i}{}_{j} = - {\cal Q}_{\mu\, j}{}^i\;,\qquad
{\cal Q}_{\mu i}{}^i=0\;,\qquad
{\cal P}_\mu{}^{ijkl} = \ft1{24}\,\varepsilon^{ijklmnpq}\, {\cal P}_{\mu\,mnpq}
\;,
\eea
as a consequence of ${\cal V}_M{}^{\underline N}$ being related to
an $E_{7(7)}$ element by multiplication with a constant matrix.
The integrability conditions of (\ref{defQP}) yield the
Cartan-Maurer equations,
\bea
  \label{eq:GECM-Q-P}
{\cal F}({\cal Q})_{\mu\nu\,i}{}^j ~\equiv~  2\partial_{[\mu} \mathcal{Q}_{\nu] i}{}^j
 + \mathcal{Q}_{[\mu i}{}^k\,
\mathcal{Q}_{\nu]k}{}^j
 & = &  \ft43\,{\cal P}_{[\mu}{}^{\!jklm}
  \, {\cal P}_{\nu]iklm} -\ft23 i\,\mathcal{H}_{\mu\nu}^{M} \,
  (\check{X}_{M}){}^{PQ}\, {\cal V}_{P\,ik} {\cal V}_Q{}^{jk}
  \nonumber \\[1ex]
   {\cal D}_{[ \mu}{\cal P}_{\nu]}{}^{ijkl} &=& - \ft12
   i \,\mathcal{H}_{\mu\nu}^{M} \,
   (\check{X}_{M}){}^{PQ} \,{\cal V}_{P}{}^{ij}  {\cal V}_{Q}{}^{kl}  \,,
   \label{Cartan-Maurer}
\eea
with the $SU(8)$ covariant derivative ${\cal D}_{\mu}$
and the covariant field strength $\mathcal{H}_{\mu\nu}^{M}$ from (\ref{defH}).
Note that its part carrying the two-forms $B_{\mu\nu\,\alpha}$ drops from
(\ref{Cartan-Maurer}) due to the orthogonality relation (\ref{QQZ}).

%
\mathon
\subsection{The $T$-tensor}
\mathoff
%
Following~\cite{deWit:1982ig,deWit:2007mt} we define the $T$-tensor
as the gauge group generator (\ref{defX}) dressed with the scalar vielbein
\bea
(T_{ij})^{klmn} &\equiv&
\ft12\,({\cal V}^{-1})_{ij}{}^M ({\cal V}^{-1})^{kl}{}^N\,(X_{M})_{N}{}^K\,{\cal V}_K{}^{mn}
\;,\qquad
\mbox{etc.}
\;.
\label{TX}
\eea
The various components of this tensor will show up in the modified field equations
of the gauged theory and parametrize the new couplings.
The linear constraints (\ref{linear}) satisfied by the embedding tensor can be made explicit by parametrizing
the $T$-tensor in terms of the irreducible $SU(8)$ tensors $A^{ij}$, $A_i{}^{jkl}$, $B^{ij}$, transforming in the
${\bf 36}$, ${\bf 420}$, and the ${\bf 28}$, respectively,\footnote{
I.e.\ $A^{ij}=A^{(ij)}$, $A_i{}^{jkl}=A_i{}^{[jkl]}$, $A_i{}^{ikl}=0$, $B^{ij}=B^{[ij]}$,
and complex conjugates $(A^{ij})^*=A_{ij}$, etc.
}
according to
\beqs
(T_{ij})_{kl}{}^{mn} &=&  \ft12\delta_{[k}^{[m} A^{n]}{}_{l]ij} +  \delta^{mn}_{[i[k} A_{l]j]} - \ft16 (8\,\delta^{mn}_{[i[k} B_{l]j]} +  \delta^{mn}_{kl} B_{ij}) - \ft12\delta^{mn}_{kl} B_{ij}\;,\nonumber\\[.5ex]
(T_{ij})^{rs}{}_{pq} &=& - \ft12\delta_{[p}^{[r} A^{s]}{}_{q]ij} -  \delta^{rs}_{[i[p} A_{q]j]} + \ft16 (8\,\delta^{rs}_{[i[p} B_{q]j]} + \delta^{rs}_{pq} B_{ij}) - \ft12\delta^{rs}_{pq} B_{ij}\;,\nonumber\\[.5ex]
(T_{ij})_{kl \, pq} &=& \ft{1}{24} \epsilon_{klpqrstu} \delta^r_{[i} A_{j]}{}^{stu}
+ \ft1{12} \epsilon_{klpqijtu} B^{tu}\;,\nonumber\\[.5ex]
(T_{ij})^{rs \; mn} &=& \delta^{[r}_{[i} A_{j]}{}^{smn]} + 2 \delta_{ij}^{[rs} B^{mn]}
\;.
\label{TAB}
\eeqs
The tensors $A^{ij}$, $A_i{}^{jkl}$
together with their complex conjugates fill the ${\bf 912}$
representation~$\Theta_M{}^\alpha$ of
the embedding tensor and carry the structure of the standard gaugings.
The tensor $B^{ij}$ is related to the new components $\vartheta_M$
of the embedding tensor according to
\be
\vartheta_M = {\cal V}_M{}^{ij} B_{ij} + {\cal V}_{M\; ij} B^{ij}
\;,
\label{defB}
\ee
and contains all the new contributions due to the gauging of the trombone generator.
Together, the tensors $A$ and $B$ will describe the scalar couplings of the gauged theory.
From their definition (\ref{TAB}) and (\ref{TX}) one derives the differential relations
\beqs
{\cal D}_{\mu} A{}^{ij} &=& \ft{1}{3} {A{}^{(i}}_{klm} {\cal P}_{\mu}{}^{j)klm}\;,\\
{\cal D}_{\mu} A_{i}{}^{jkl} &=&2A_{im}  {\cal P}_{\mu}{}^{mjkl} + 3 {\cal P}_{\mu}{}^{mn[jk} A^{l]}{}_{imn}
+ {\cal P}_{\mu}{}^{mnp[j}\delta^k_i A^{l]}{}_{mnp}
 \;,\\
{\cal D}_{\mu} B_{ij} &=& - {\cal P}_{\mu\; ijkl} B^{kl}\;,
\eeqs
where again ${\cal D}_{\mu}$ refers to the $SU(8)$ covariant derivative with the composite connection
${\cal Q}_{\mu\,i}{}^j$ from (\ref{defQP}).

For the supersymmetry algebra it will also be useful to compute the tensor
${Z}^M{}_{KL}$ upon dressing with $({\cal V}^{-1})_{kj}{}^K  ({\cal V}^{-1})^{ij}{}^L$:
\beqs
{{Z}^M{}}_{kj}{}^{ij} &=& -\frac{3}{2} ({\cal V}^{-1\, in \, M} A_{nk} + {{\cal V}^{-1}{}}_{kl}{}^M A^{ni}) + \frac{3}{4} ({\cal V}^{-1\, mn \, M} A^i{}_{kmn} + {{\cal V}^{-1}{}}_{mn}{}^M A_k{}^{imn})\nonumber\\
&+& 4 ({\cal V}^{-1\, in \, M} B_{nk} + {{\cal V}^{-1}{}}_{kl}{}^M B^{ni}) + \frac{1}{2} \delta_k^i ({\cal V}^{-1\, mn \, M} B_{mn} + {{\cal V}^{-1}{}}_{mn}{}^M B^{mn})\;.
\nonumber\\
\label{ZM}
\eeqs

Dressing the quadratic constraints (\ref{Q1})--(\ref{Q3}) (or alternatively (\ref{XXX}))
with the scalar vielbein (\ref{VV}) induces a plethora of relations bilinear in the tensors $A$, $B$.
In appendix~\ref{app:Tids}, we have collected a number of such identities which are important in
the subsequent calculations.
Here, we only give two examples of such identities. A linear combination of the constraints
(\ref{constraints63}) transforming in the ${\bf 63}$ of $SU(8)$
shows that the traceless part of the hermitean tensor defined by
\bea
\Pi^i{}_j  &\equiv&
6A^{ik}A_{jk}-\ft13A^i{}_{mnk} A_j{}^{mnk}
+\ft43\left(A_j{}^{imn}B_{mn}+A^i{}_{jmn}B^{mn}\right)
-\ft{256}9 \, B^{ik}B_{jk}
\;,
\nonumber
\eea
vanishes
\bea
\Pi^i{}_j  &=& \ft18\delta^i_j \,\Pi^k{}_k
\;.
\label{QCcomb1}
\eea
Another useful identity is given by the self-duality equation
\bea
\Pi_{ijkl} &=& \ft1{24} \epsilon_{ijklmnpq}\,\Pi^{mnpq}
\;,\nonumber\\[1ex]
\mbox{for}\quad
\Pi_{ijkl}&\equiv&
  A^m{}_{[ijk} A_{l]m}
  -\ft34  A^m{}_{p[ij}A^p{}_{kl]m}
  +2 A^m{}_{[ijk} B_{l]m}
  -8 B_{[ij}B_{kl]}
  \;,
\label{QCcomb2}
\eea
which is obtained as a linear combination
of the constraints (\ref{constraints70}) transforming in the ${\bf 70}$ of $SU(8)$

%
\mathon
\subsection{Vector fields}
\mathoff
%

As mentioned above, only half of the 56 vector fields ${\cal A}_\mu^M$ enter the Lagrangian of the ungauged theory. This corresponds to selecting a symplectic frame, such that the vector fields split according to $\{{\cal A}_\mu^M\} \rightarrow \{{\cal A}_\mu^\Lambda, {\cal A}_{\mu\,\Lambda}\}$ into electric and magnetic fields.
Accordingly, we define the electric field strengths
${\cal H}_{\mu\nu}^\Lambda$ via (\ref{defH}) as the curvature of ${\cal A}_\mu^\Lambda$ while their magnetic duals are defined as functions of the electric vector fields according to
\bea
{\cal G}_{\mu\nu\,\Lambda}^+ &\equiv&
{\cal N}_{\Lambda\Sigma}\,{\cal H}_{\mu\nu}^{+\;\Sigma} ~+ \mbox{fermions}
\;,
\label{defGF}
\eea
with the complex matrix ${\cal N}_{\Lambda\Sigma}$ defined by
$\,{\cal V}^{\Sigma\, ij}\,{\cal N}_{\Lambda\Sigma}\equiv - {\cal V}_\Lambda{}^{ij}$\,,
and where the superscript~$^\pm$ refers to the (anti-)selfdual part of the
field strength.
The fermionic part of (\ref{defGF}) is explicitly given in~\cite{Cremmer:1979up,deWit:1982ig,deWit:2007mt}.
We define the full symplectic vector ${\cal G}_{\mu\nu}^M \equiv
({\cal H}_{\mu\nu}^\Lambda, {\cal G}_{\mu\nu\,\Lambda})$\,, which will
in particular enter the fermionic field equations and supersymmetry transformation rules.
By construction, it allows the decomposition
\bea
{\cal G}_{\mu\nu}^M &=&
({\cal V}^{-1})^{ij\,M} {\cal G}_{\mu\nu\,ij}^{+}
+
({\cal V}^{-1})_{ij}{}^M {\cal G}_{\mu\nu}^{-\;ij}
~+ \mbox{fermions}
\;,
\eea
into its selfdual and anti-selfdual part.
In contrast, we introduce the field strengths ${\cal H}_{\mu\nu}^{ij}$ and
${\cal H}_{\mu\nu\,ij}$ as the dressed version
\bea
{\cal H}_{\mu\nu}^M &=&
({\cal V}^{-1})^{ij\,M} {\cal H}_{\mu\nu\,ij}
+
({\cal V}^{-1})_{ij}{}^M {\cal H}_{\mu\nu}^{\;ij}
\;,
\eea
of the covariant non-abelian field strengths introduced in (\ref{defH}),
that combine electric and magnetic vector fields.
Note that ${\cal H}_{\mu\nu}^\Lambda={\cal G}_{\mu\nu}^\Lambda$ is identically
satisfied, whereas
${\cal H}_{\mu\nu\,\Lambda}={\cal G}_{\mu\nu\,\Lambda}$ describes the
first order duality relation between electric and magnetic vector fields.

%
\section{Supersymmetry algebra}
\label{sec:susy}

Before deriving the full set of supersymmetric equations of motion,
we establish the supersymmetry transformation rules by verifying the supersymmetry algebra.
Under supersymmetry, the bosonic fields transform as
\bea
  \delta_{\epsilon} e_{\mu}{}^{a}&=&
  \bar\epsilon^{i}\gamma^{a}\psi_{\mu i} ~+~
  \bar\epsilon_{i}\gamma^{a}\psi_{\mu}{}^i \;, \nonumber\\[1ex]
  \delta_{\epsilon}{\cal V}_M{}^{ij} &=&  2\sqrt{2}\,{\cal V}_{M kl} \, \Big(
  \bar\epsilon^{[i}\chi^{jkl]}+\ft1{24}\varepsilon^{ijklmnpq}\,
  \bar\epsilon_{m}\chi_{npq}\Big)   \,,  \nonumber \\[1ex]
    \delta_{\epsilon} {\cal A}_{\mu}^{M}
    &=&
    -i\,\Omega^{MN} {\cal V}_N{}^{ij}\,\Big(
    \bar\epsilon^{k}\,\gamma_{\mu}\,\chi_{ijk}
    +2\sqrt{2}\, \bar\epsilon_{i}\,\psi_{\mu j}\Big)~+~ {\rm c.c.}
    \;, \nonumber \\[1ex]
    \delta_{\epsilon} B_{\mu\nu\,\alpha} &=&\ft{2}{3} \sqrt{2} \,
    (t_{\alpha})^{PQ}\, \Big( {\cal V}_{P\,ij} {\cal
    V}_{Q\,kl}\,
    \bar\epsilon^{[i}\,\gamma_{\mu\nu}\,\chi^{jkl]}
    + 2 \sqrt{2}\, {\cal V}_{P\,jk} {\cal V}_{Q}{}^{ik}\,
    \bar\epsilon_{i}\,\gamma_{[\mu}\,\psi_{\nu]}{}^{j}
    ~+~ {\rm c.c.}\Big)  \nonumber\\
    &&{}
    +2(t_{\alpha})_{MN}\,{\cal A}_{[\mu}^{M}\,\delta
    {\cal A}_{\nu]}^{N} \;.
    \label{susybosons}
\eea
while the transformation of the fermions is given by
\beqs
\delta_{\epsilon} \psi_{\mu}^i &=& 2 {\cal D}_{\mu} \epsilon^i + \frac{\sqrt{2}}{4} {{\cal G}}^-_{\rho \sigma}{}^{ij} \gamma^{\rho \sigma} \gamma_{\mu} \epsilon_j + \sqrt{2} A^{ij} \gamma_{\mu} \epsilon_j -2\sqrt{2}  B^{ij} \gamma_{\mu} \epsilon_j\;,\nonumber\\
\delta_{\epsilon} \chi^{ijk} &=& -2 \sqrt{2} {{\cal P}}_{\mu}{}^{ijkl} \gamma^{\mu} \epsilon_l + \frac{3}{2} {{\cal G}}^-_{\mu \nu}{}^{[ij} \gamma^{\mu \nu} \epsilon^{k]} -2 A_l{}^{ijk} \epsilon^l -4 B^{[ij} \epsilon^{k]}\;,
\label{susyfermions}
\eeqs
up to higher order fermion terms.
Except for the respective last terms in the fermionic transformation rules
(carrying the tensor $B_{ij}$), these supersymmetry transformations
are known from~\cite{deWit:1982ig,deWit:2007mt}. The structure of the new terms follows from the $SU(8)$ representation content, their factors are determined from the closure of the supersymmetry algebra.
This algebra is given by
\begin{equation}
  \label{eq:susy-algebra1}
  {}[\delta(\epsilon_1),\delta(\epsilon_2)] = \xi^\mu \hat D_\mu +
  \delta_{\rm Lor}(\Omega^{ab}) + \delta_{\rm susy}(\epsilon_3) +
  \delta_{\rm SU(8)}(\Lambda^i{}_j) +  \delta_{\rm gauge}(\Lambda^M) +
  \delta_{\rm gauge}(\Xi_{\mu \alpha})  \,.
\end{equation}
The first term refers to a covariantized general coordinate transformation
with diffeomorphism parameter
\be
\xi^\mu = 2\, \bar\epsilon_2{}^i \gamma^\mu \epsilon_{1\, i} + 2\, \bar\epsilon_{2\; i} \gamma^\mu \epsilon_1{}^i
\;,
\ee
and including terms of order $g$ induced by the gauging.
The last two terms refer to
gauge transformations (\ref{gaugeV1}), (\ref{gaugeV2}) and (\ref{gaugeT}),
with parameters
\bea
\Lambda^N &=& -4 i \sqrt{2} \; \Omega^{NP} \left({\cal V}_P{}^{mn} \bar\epsilon_{2\, m} \epsilon_{1\, n} - {\cal V}_{P\;mn} \bar\epsilon_2^m \epsilon_1^n\right)
\;,
\nonumber\\
  \Xi_{\mu\alpha} &=& -\ft83 (t_\alpha)^{PQ}\,
  {\cal V}_{P\,ik} {\cal V}_Q{}^{jk} \left(\bar\epsilon_2{}^i \gamma_\mu \epsilon_{1j}
  + \bar\epsilon_{2j} \gamma_\mu \epsilon_1{}^i\right)
\;,
\label{LX}
\eea
respectively. Up to the contributions from the new terms in the supersymmetry transformation rules,
the supersymmetry algebra
has been verified in \cite{deWit:1982ig,deWit:2007mt}. In presence of
$B_{ij}$, $B^{ij}$, the commutator (\ref{eq:susy-algebra1}) evaluated on the vielbein acquires the additional terms
\bea
[\delta_{\epsilon_1}, \delta_{\epsilon_2}] \, e_{\mu}{}^a &=&
\dots \quad
-4 \sqrt{2}\, \left(B^{mn} \bar\epsilon_{2\, m} \epsilon_{1\, n} + B_{mn} \bar\epsilon_2^m \epsilon_1^n \right) e_{\mu}{}^a
\;.
\eea
These precisely reproduce the action
of a scaling gauge transformation with parameter (\ref{LX}) on the vielbein
\beqs
\delta_{\Lambda} e_{\mu}{}^a &=&  \Lambda^M \vartheta_M\,  {\bf t}_0 \cdot e_{\mu}{}^a \\ \nn
&=& -4 i \sqrt{2} \; \Omega^{MN} \vartheta_M ({\cal V}_N{}^{mn} \bar\epsilon_{2\, m} \epsilon_{1\, n} - {\cal V}_{N\;mn} \bar\epsilon_2^m \epsilon_1^n) \, e_{\mu}{}^a\\ \nn
&=&
-4 \sqrt{2}\, \left(B^{mn} \bar\epsilon_{2\; m} \epsilon_{1\; n} + B_{mn} \bar\epsilon_2^m \epsilon_1^n \right) e_{\mu}{}^a
\;,
\eeqs
where we have used (\ref{eq:VV-orthogonal}) and (\ref{defB}).
Similarly, one may check that the terms carrying the scalar tensors (\ref{TAB})  in the supersymmetry commutator
on the scalar fields combine into
\beqs\nn
{\cal V}^{-1\, ij \; M} \left[\delta_{\epsilon_1}, \delta_{\epsilon_2}\right] {\cal V}_M{}^{kl} &=&
\dots\quad -8\sqrt{2}\left(
T^{mn}{}^{ijkl} \,\bar\epsilon_{2\, m} \epsilon_{1\, n}+
T_{mn}{}^{ijkl} \, \bar\epsilon_2^m \epsilon_{1}^n \right)
\nonumber\\[.5ex]
&=& \dots\quad + {\cal V}^{-1\;ij \;M}\; \Lambda^N (X_N)_M{}^K\, {\cal V}_K{}^{kl}
\;,
\qquad
\mbox{etc.}
\nonumber
\eeqs
and consistently reproduce the action
of a gauge transformation with parameter (\ref{LX}).
In checking the supersymmetry algebra on the vielbein and the scalar fields,
we have fixed all the new factors in the supersymmetry transformation rules (\ref{susyfermions}).
As a consistency check, one may further verify  that the algebra also closes
on the vector and the tensor gauge fields.

%
\section{Equations of motion}
\label{sec:eom}

%
\subsection{Einstein equations}
%

Having established the supersymmetry algebra, we can now determine the deformed equations
of motion by requiring covariance under the new supersymmetry transformation rules.
As there is no longer an action underlying the gauged theory, we have to work directly on the level
of the equations of motion.
This derivation of the supersymmetric field equations is based on reference~\cite{LeDiffon:2010}.
We will start from the gravitino equations of motion for which we use the following ansatz
\beqs\nn
0~=~ ({\cal E}_{\psi})^\mu{}_i &\equiv&
-e^{-1} \varepsilon^{\mu \nu \rho \sigma} \gamma_{\nu} {\cal D}_\rho \psi_{\sigma\, i}
- \ft{\sqrt{2}}{6}  \gamma^\nu \gamma^\mu  \chi^{jkl} \, {\cal P}_{\nu \, jkli}
- \ft{\sqrt{2}}{4}  {\cal G}^{+\; \rho \sigma}{}_{ij} \gamma^{[\mu} \gamma_{\rho \sigma} \gamma^{\nu]} \psi_{\nu}{}^j
\nonumber\\[.5ex]
&& {}
+ \ft{1}{8}  {\cal G}^{-\; \rho \sigma}{}^{jk} \gamma_{\rho \sigma} \gamma^{\mu} \chi_{jki}
+ \sqrt{2}  A_{ij} \gamma^{\mu \nu} \psi_{\nu}{}^j + \ft{1}{6}   A_{i}{}^{jkl} \gamma^{\mu} \chi_{jkl}
\nonumber\\[.5ex]
&& {}
+\lambda\,  \sqrt{2}  B_{ij} \gamma^{\mu \nu} \psi_{\nu}{}^j + \zeta \,  B^{kl} \gamma^{\mu} \chi_{ikl}
\;.
\label{eomgrav}
\eeqs
Except for the last two terms, these are the equations obtained from variation of the Lagrangian~\cite{deWit:1982ig,deWit:2007mt} of the gauged theory. While the $SU(8)$ structure of these two additional terms is fully determined by the representation content, we will in the following determine their unknown coefficients $\lambda$ and $\zeta$ by compatibility with supersymmetry.

E.g.\ vanishing of the ${\cal D}_\mu \epsilon$ terms in the supersymmetry variation of (\ref{eomgrav}) imposes
\bea
2\sqrt{2}\,B_{ij}  \left( \lambda \gamma^{\mu\nu} - e^{-1} \varepsilon^{\mu\nu\rho\sigma} \gamma_{\rho\sigma}    \right)
{\cal D}_{\nu} \epsilon^{j}
 &=& 0\;,
\eea
from which we deduce $\lambda=-2$.
Vanishing of the terms linear in $B {\cal G}^\pm \epsilon$ further
determines $\zeta=-5/3$, but we will for the moment keep the parameters in the formulas
so as to allow for further consistency checks.

Let us concentrate on the part of the supersymmetry variation of (\ref{eomgrav}) which is bilinear in the scalar tensors $A$, $B$
which originate from variation of its last four terms. We obtain
\bea
&&{}
\Big\{
6 A_{ik} A{}^{jk} - \frac{1}{3} A_{i}{}^{klm} {A{}^j}_{klm} + 12 A_{ik} B^{jk}
+ 6 \lambda B_{ik} A^{jk}\label{bil1}\\
&&{}\qquad
- \frac{2}{3} A_{i}{}^{jlm} B_{lm} - 2 \zeta {A{}^j}_{ikl} B^{kl} +( 12\lambda + \frac{8 \zeta}{3})\, B_{ik} B^{jk}
 - \frac{4\zeta}{3} \delta_i^j B_{kl} B^{kl}
\Big\}\;\;  \times\;\; \gamma^\mu \epsilon_j
\;.
\nonumber
\eea
Only the $(ij)$-trace of the braced expression can be absorbed into a modification of the Einstein equations. In particular
its anti-hermitean part must vanish. Indeed, this follows from the first of the bilinear constraint relations (\ref{constraints63})
provided that $\lambda=2\zeta+\frac43$, which is satisfied for our above choice of constants.
With this value for $\lambda$, the expression (\ref{bil1}) reduces to its hermitean part
\bea
&&{}
\Big\{
6 A_{ik} A{}^{jk} - \frac{1}{3} A_{i}{}^{klm} {A{}^j}_{klm} + (10+6\zeta)( A_{ik} B^{jk}+B_{ik} A^{jk})\;\;\;
\\
&&{}\qquad
-(\ft13+\zeta) (A_{i}{}^{jlm} B_{lm} + {A{}^j}_{ikl} B^{kl}) +( 16 + \frac{80 \zeta}{3})\, B_{ik} B^{jk}
 - \frac{4\zeta}{3} \delta_i^j B_{kl} B^{kl}
\Big\}\; \times\; \gamma^\mu \epsilon_j
\;.
\nonumber
\eea
Finally, we observe that with the above value $\zeta=-5/3$ all coefficients precisely reproduce
the linear combination appearing in the quadratic constraint (\ref{QCcomb1}), such that the
full expression reduces to its trace part
\bea
 \Big\{\frac{3}{4} A_{kl} A{}^{kl} - \frac{1}{24} A_{n}{}^{jkl} {A{}^n}_{jkl}
- \frac{4}{3} B_{kl} B^{kl}\Big\} \;  \times\;\; \gamma^\mu \epsilon_i
\;,
\eea
which can be absorbed into the modified Einstein equations.
Another important ingredient in the calculation is the evaluation of the commutator
\bea{}
\gamma^{\mu\nu\rho}\,[{\cal D}_\nu, {\cal D}_\rho ] \,\epsilon_i &=&
\gamma^{\mu\nu\rho}\,\left(
\ft12{\cal F}({\cal Q})_{\nu\rho\,i}{}^j\,\epsilon_j
-\ft14 \widehat{\cal R}_{\nu\rho}{}^{ab}\gamma_{ab}\,\epsilon_i
-\ft12\vartheta_M {\cal H}_{\nu\rho}^M\,\epsilon_i
\right)
\nonumber\\
&=&
\left(\ft12g^{\mu\nu}\widehat{{\cal R}}-\widehat{{\cal R}}^{(\mu \nu )}\right)
\gamma_\nu \epsilon_i
+\vartheta_M \left({\cal H}^{-\,\mu\nu}{}^M-3{\cal H}^{+\,\mu\nu}{}^M\right)
\gamma_{\nu}\epsilon_i
\nonumber\\
&&{}
+\ft12
\gamma^{\mu\nu\rho}\,
{\cal F}({\cal Q})_{\nu\rho\,i}{}^j\epsilon_j
\;,
\eea
with the modified Riemann tensor and the curvature of
the $SU(8)$ connection from (\ref{modRR}) and (\ref{Cartan-Maurer}), respectively

Putting all this together,
a somewhat lengthy calculation shows that the full supersymmetry
variation of the Rarita-Schwinger equation (\ref{eomgrav}) eventually takes the form
\bea
\delta_\epsilon ({\cal E}_{\psi})^\mu{}_i &=&
({\cal E}_{\rm Einstein})^{\mu\nu}\,\gamma_\nu \epsilon_i
-2\sqrt{2} \,({\cal E}_{\rm vector})^{\mu}{}_{ij} \, \epsilon^j
\;,
\label{var32}
\eea
where $({\cal E}_{\rm Einstein})^{\mu\nu}$ and $({\cal E}_{\rm vector})^{\mu}{}_{ij}$
denote the modified Einstein and the vector field equations of motion, respectively.
In bringing the supersymmetry variation into this form, we have in particular made use
of the equations
\bea
X_M\left( {\cal G}_{\mu\nu}^M - {\cal H}_{\mu\nu}^M \right)
~=~ 0 ~=~
\vartheta_M \left( {\cal G}_{\mu\nu}^M - {\cal H}_{\mu\nu}^M \right)
\;.
\label{eom:1order}
\eea
For a purely electric gauging, these equations are identically satisfied. In presence
of magnetic charges, these equations represent the first order duality equations
between electric and magnetic vector fields.
The second order field equations obtained in (\ref{var32}) read
\bea
({\cal E}_{\rm Einstein})^{\mu\nu} &=&
\widehat{{\cal R}}^{(\mu \nu )}-\ft12g^{\mu\nu}\widehat{{\cal R}}
+\ft16\,{\cal P}^{(\mu}_{ijkl} {\cal P}^{\nu)ijkl}
-\ft1{12} g^{\mu\nu}\,{\cal P}^\rho_{ijkl} {\cal P}_\rho^{ijkl}
-{\cal G}^+{\!}_\rho{}^{(\mu}{}_{jk} \,{\cal G}^{\nu)\rho-\,jk}
\nonumber\\[1ex]
&&{}
+g^{\mu\nu} \left( \ft{3}{4} A_{kl} A{}^{kl} - \ft{1}{24} A_{n}{}^{jkl} {A{}^n}_{jkl}
- \ft{4}{3} B_{kl} B^{kl} \right)
\;.
\label{eom:Einstein}
\eea
and
\bea
({\cal E}_{\rm vector})^{\mu}{}_{ij} &=&
{\cal D}_\nu {\cal G}^{+\nu\mu}{}_{ij} +
{\cal P}_{\nu\,ijkl}\,{\cal G}^{-\,\nu\mu\,kl}
-\ft13
\left(
A_{[i}{}^{nkl} {\cal P}^\mu{}_{j]nkl}+
4 B^{kl}{\cal P}^\mu{}_{ijkl}
\right)
\;.
\label{eom:vector}
\eea
The modified Einstein equations show that the presence of the tensor $B_{ij}$ induces a positive contribution to the effective cosmological constant. This is a typical feature of the theories with trombone gauging~\cite{Howe:1997qt,Lavrinenko:1997qa,Chamblin:2001dx,Bergshoeff:2002nv,Kerimo:2003am,Kerimo:2004md,LeDiffon:2008sh}.

%
\subsection{Scalar field equations}
%

We start from the following ansatz for the equations of motion for the
spin-$1/2$ fermion fields $\chi_{ijk}$
\beqs
0 ~=~ ({\cal E}_\chi)_{ijk} &\equiv& - \ft{1}{6}  \gamma^{\mu} {\cal D}_{\mu} \chi_{ijk}
- \ft{\sqrt{2}}{6}  \gamma^{\nu} \gamma^{\mu} \psi_{\nu}{}^l {\cal P}_{\mu\; ijkl}
+ \ft{1}{8}  \gamma_\rho\gamma_{\mu\nu} \psi^{\rho}{}_{[k} {\cal G}^{+ \mu\nu}{}_{ij]}
\nonumber\\[.5ex]
&& {}
+ \ft{\sqrt{2}}{288}  \epsilon_{ijklmnpq} \gamma^{\mu\nu} \chi^{lmn} {\cal G}^{-}{}_{\mu\nu}{}^{pq}
- \ft{1}{6}   {A{}^l}{}_{ijk} \gamma^{\mu} \psi_{\mu\; l}
+ 2  A_{ijk,\, lmn} \chi^{lmn}
\nonumber\\[.5ex]
&& {}
-\ft13  \gamma^{\mu} \psi_{\mu[k}  B_{ij]} + \ft{\sqrt{2}}{36}  \epsilon_{ijklmnpq} \chi^{lmn} B^{pq}
\;,
\label{eom:fermions}
\eeqs
with the scalar tensor
\bea
  A{}_{ijk,lmn} = \ft1{144} \sqrt{2}\,
  \epsilon_{ijkpqr[lm}\,A_{n]}{}^{pqr}
  \;.
\eea
Again, up to the last two terms whose structure is determined by $SU(8)$,
equations (\ref{eom:fermions}) follow from varying
the Lagrangian of~\cite{deWit:1982ig,deWit:2007mt}. The new coefficients are
fixed by compatibility with supersymmetry and follow as in the last section
by imposing the vanishing of the linear terms of the form $B {\cal D} \epsilon$
and $B {\cal P} \epsilon$ in the supersymmetry
variation of (\ref{eom:fermions}).

Again we first focus on the part of the supersymmetry variation of $({\cal E}_\chi)_{ijk}$
which is bilinear in the scalar tensors $A$, $B$ and find
\bea
&&-\; \frac{\sqrt{2}}{3} \;
\Big\{
2{A{}^r}{}_{ijk} A_{lr}
+4 {A{}^r}{}_{ijk} B_{lr}
+4 B_{[ij} A_{k]l}
-8 B_{[ij} B_{k]l}
-\frac{1}{9} \epsilon_{ijklrmnp} A_{q}{}^{rmn}B^{pq}
\nonumber\\[.5ex]
&&{}
\qquad\qquad
+\frac{1}{12} \epsilon_{ijkrmnpq} A_{l}{}^{pqs}
(A_{s}{}^{rmn} +\ft83 \delta_s^r B^{mn} )
+\frac13 \epsilon_{ijklmnpq} B^{mn} B^{pq}
 \Big\} \;\times\;
 \epsilon^l
 \;.
 \qquad\quad
\eea
Upon adding a proper linear combination of the two quadratic constraints (\ref{constraints378})
in the ${\bf 378}$, this expression reduces to
\bea
&&-\; \frac{\sqrt{2}}{3} \;
\Big\{
2{A{}^r}{}_{[ijk} A_{l]r}
+4 {A{}^r}{}_{[ijk} B_{l]r}
-8 B_{[ij} B_{kl]}
\nonumber\\[.5ex]
&&{}
\qquad\qquad\qquad\qquad\qquad
+\frac{1}{16} \epsilon_{ijklmnpq} \left(A_{r}{}^{pqs} A_{s}{}^{mnr}
+\ft{16}3 B^{mn} B^{pq}\right)
 \Big\} \;\times\;
 \epsilon^l
 \;,
 \qquad\quad
\eea
which is manifestly antisymmetric in $[ijkl]$\,. Finally, the combination of quadratic constraints (\ref{QCcomb2}) can be
used to simplify this expression to the manifestly self-dual expression
\bea
&&-\; \frac{\sqrt{2}}{3} \;
\Big\{\,
{\cal C}_{ijkl} + \ft1{24} \epsilon_{ijklmnpq} {\cal C}^{mnpq}\,
 \Big\} \;\times\;
 \epsilon^l
 \;,
 \qquad\quad
\eea
with the tensor
\bea
{\cal C}_{ijkl}&=&
{A{}^m}_{[ijk} A_{l]m}
+ \ft34 {A{}^m}_{n[ij} {A^n}_{kl]m}
+ 2 {A{}^m}_{[ijk} B_{l]m}
\;.
\label{defC}
\eea
This expression will be part of the modified scalar field equations.
After some more calculation, and using the first order field equations
(\ref{eom:1order}), the
full supersymmetry variation of the fermionic field equation $({\cal E}_\chi)_{ijk}$
eventually takes the form
\bea
\delta_\epsilon ({\cal E}_\chi)_{ijk} &=&
\ft{\sqrt{2}}3 \,({\cal E}_{\rm scalars})_{ijkl}\,\epsilon^l
-\gamma_\mu \,({\cal E}_{\rm vector})^{\mu}{}_{[ij}\,\epsilon_{k]}
\;.
\eea
with the vector field equations from (\ref{eom:vector}) and the full scalar field equations
given by
\bea
({\cal E}_{\rm scalars})_{ijkl} &=&
{\cal D}^\mu{\cal P}_{\mu\,ijkl} -
\ft32
{\cal G}^+_{\mu\nu\,[ij}{\cal G}^{+\,\mu\nu}{}_{kl]}
-\ft1{16}\epsilon_{ijklpqrs}
{\cal G}^{-\mu\nu\,pq}{\cal G}^{-}{}_{\mu\nu}{}^{rs}
\nonumber\\
&&{}
-{\cal C}_{ijkl}-\ft1{24}\epsilon_{ijklpqrs}\, {\cal C}^{pqrs}
\;.
\label{eom:scalars}
\eea
We note that the term bilinear in the tensor $B_{ij}$ has dropped out from the
scalar field equations. Also this is a characteristic feature for theories with trombone gauging.\footnote{
Let us correct a misprint in reference \cite{LeDiffon:2008sh}:
the last term in equation (4.44) of that reference is in fact absent in accordance with the equation obtained
by dimensional reduction of (\ref{eom:scalars}).
}
 As a consequence, for pure trombone gaugings
($\Theta_M{}^\alpha=0$, implying that $A_{ij}=0=A_i{}^{jkl}$)
 a simple solution to the bosonic field equations is given by
a de Sitter geometry with all scalar and vector fields vanishing.
We shall discuss this solution in more detail below.
Let us finally note, that due to the presence of the mixed term ${A{}^m}_{[ijk} B_{l]m}$ the scalar field
equations (\ref{eom:scalars}) cannot be integrated up to a scalar potential.

%
\subsection{Yang-Mills equations}
%

The vector field equations of motion (\ref{eom:vector}) can be rewritten equivalently as
\bea
{\cal V}_M{}^{ij} \, {\cal D}_{[\mu}\,{\cal G}_{\nu\rho]}{}^{M} &=&
-\ft19e\varepsilon_{\mu\nu\rho\sigma}\,
\Big\{
A^{[i}{}_{nkl} {\cal P}^{\sigma\,j]nkl}+
4 B_{kl}{\cal P}^{\sigma\,ijkl}
\Big\}
\;,
\label{eom:vector0}
\eea
or
\bea
{\cal D}_{[\mu}\,{\cal G}_{\nu\rho]}{}^{M} &=& \ft19 e\varepsilon_{\mu\nu\rho\sigma}
Z^{M\alpha} (t_{\alpha})^{KL} {\cal V}_{K}{}^{ij}{\cal V}_{L}{}^{kl} {\cal P}^{\sigma}{}_{ijkl}
\;,
\label{eom:vector1}
\eea
with the tensor $Z^{M\alpha}$ from (\ref{defZZ}).
Since in the derivation of the field equations we have also come across the first order
duality equations (\ref{eom:1order}) for the vector fields, an immediate question is the
compatibility of these equations with the second order field equations. Upon contracting
equation (\ref{eom:vector1}) with $\vartheta_M$ or $X_M$, its r.h.s.\ vanishes by virtue of (\ref{QQZ})
while on the l.h.s.\ the first order duality equations (\ref{eom:1order}) allow to replace
${\cal G}_{\nu\rho}^{M}$ by the covariant field strength ${\cal H}_{\nu\rho}^{M}$. Then also
the l.h.s.\ vanishes by virtue of the Bianchi identity (\ref{Bianchi}) and (\ref{QQZ}).
Both sets of equations are thus compatible.
As another consistency check, we observe that
upon applying the operator $\varepsilon^{\mu\nu\rho\tau} {\cal D}_\tau$,
the l.h.s.\ of (\ref{eom:vector1}) reduces to
\bea
 - \ft12
 \varepsilon^{\mu\nu\rho\tau} \,
 {\cal G}^K_{\mu\nu} {\cal G}_{\rho\tau}^{L} \, X_{KL}{}^M
 &=&
 - Z^{M\alpha}\,(t_{\alpha})_{KL}\,
 \left(
 {\cal G}^{+\;K}_{\mu\nu} {\cal G}^{+\mu\nu\;L}
 +{\cal G}^{-\;K}_{\mu\nu} {\cal G}^{-\mu\nu\;L}
 \right)
 \;,
\eea
whereas the r.h.s.\ contains the scalar field equation (\ref{eom:scalars}).
This provides a number of important non-trivial consistency checks on the set of bosonic
field equations that we have obtained from supersymmetry variation,
but not derived from an action.

%
\subsection{Maximally symmetric solutions and mass matrices}
\label{subsec:masses}
%

According to (\ref{eom:scalars}), a solution to the field equations with constant scalar
and vanishing vector fields requires the anti-selfduality condition
\bea
{\cal C}_{ijkl}+\ft1{24}\varepsilon_{ijklpqrs}\, {\cal C}^{pqrs} &=& 0\;,
\label{extremum}
\eea
for the scalar field dependent
tensor ${\cal C}_{ijkl}$ from (\ref{defC}). For the standard gaugings, this is precisely
the condition for an extremal point of the scalar potential~\cite{deWit:1982ig}.
In presence of the local scaling symmetry, however, we recall that the scalar field equations
can in general not be integrated up to a scalar potential.
Any solution to (\ref{extremum}) yields a solution to the field equations with maximally
symmetric four-dimensional spacetime and cosmological constant
\bea
\Lambda&=&
-\ft{3}{4} A_{kl} A{}^{kl} + \ft{1}{24} A_{n}{}^{jkl} {A{}^n}_{jkl}
+ \ft{4}{3} B_{kl} B^{kl}
\;.
\label{coco}
\eea
The spectrum of the theory around this solution can be obtained by linearizing
the field equations.
Using (\ref{actionE7}), linearization of
the scalar field equations (\ref{eom:scalars})
around a solution of (\ref{extremum}) yields to lowest order
\bea
\Box\, \phi_{ijkl} &=&
{\cal M}_{ijkl}{}^{mnpq}\,\phi_{mnpq} + {\cal O}(\phi^2)
\;,
\eea
with self-dual scalar fields $\phi_{ijkl}=\ft1{24}\varepsilon_{ijklpqrs}\, \phi^{pqrs}$ and
the scalar mass matrix ${\cal M}_{ijkl}{}^{mnpq}$
whose symmetric part is given by
\bea
 {\cal M}_{ijkl}{}^{mnpq}\,\phi^{ijkl}\phi_{mnpq} &=&
6\left(A_{m}{}^{ijk}A^{l}_{ijn}\!-\!\ft14A_{i}{}^{jkl}A^{i}_{jmn}\!-\!
A_{m}{}^{ikl} B_{in}\!-\!A^{k}{}_{imn}B^{il}\right)\phi^{mnpq}\phi_{klpq}
\nonumber\\
&&{}
+\left(\ft5{24}\,A_{i}{}^{jkl}A^{i}{}_{jkl}-\ft12A_{ij}A^{ij}\right) \phi^{mnpq}\phi_{mnpq}
\nonumber\\
&&{}
-\ft23\,A_{i}{}^{jkl} A^{m}{}_{npq} \, \phi^{inpq}\phi_{jklm}
\;,
\label{Mscalar_sym}
\eea
while its antisymmetric part reads
\bea
{\cal M}_{ijkl}{}^{mnpq}\,\phi_{[1}^{ijkl}\phi^{\vphantom{j}}_{2]\,mnpq}  &=&
\ft83 \left(B_{rs}A_{i}{}^{rs[m}-B^{rs}A^{m}{}_{rs[i}\right)
\phi_{[1}^{ijkl}\phi^{\vphantom{j}}_{2]\,mjkl}
\nonumber\\
&&{}
-4 \left(A_{i}{}^{mnp} B_{jk}-A^{m}{}_{ijk} B^{np} \right)
\phi_{[1}^{ijkl}\phi^{\vphantom{j}}_{2]\,mnpl}
\;.
\label{Mscalar_alt}
\eea
The calculation of (\ref{Mscalar_sym}),  (\ref{Mscalar_alt})
makes use of identities for self-dual tensors,
such as those given in~\cite{deWit:1978sh}
as well as
of the quadratic constraints derived in appendix~\ref{app:Tids}.
For $B^{ij}=0$, the mass matrix consistently reduces to the
expression given in~\cite{deWit:1983gs}
and its antisymmetric part vanishes.

For the vector field equations (\ref{eom:vector}) we find the
linearized form
\bea
\Box\, {\cal A}_{\mu\,ij}
&=&
\ft23
\left(
A_{[i}{}^{nkl}  -
4 \delta_{[i}^n B^{kl}
\right) (T^{pq})_{j]nkl}  \, {\cal A}_\mu{\,}_{pq}
+\ft23
\left(
A_{[i}{}^{nkl}    -
4 \delta_{[i}^n B^{kl}
\right)
(T_{pq})_{j]nkl}\,{\cal A}_\mu^{\,pq}
\;,
\nonumber
\eea
from which using (\ref{TX}), (\ref{TAB}) we read off the vector mass matrix
\bea
{\cal M}_{\rm vec} &=&
\left(
\begin{array}{cc}
{\cal M}_{ij}{}^{kl} & {\cal M}_{ijkl}
\\
 {\cal M}^{ijkl}&
 {\cal M}^{ij}{}_{kl}
\end{array}
\right)
\;,
\label{M_vector}
\eea
with
\bea
{\cal M}_{ij}{}^{kl}
&=&
-\ft16 A_{[i}{}^{npq} \delta_{j]}^{[k} A^{l]}{}_{npq}
+\ft12 A_{[i}{}^{pq[k}A^{l]}{}_{j]pq}
+\ft23  \delta^{[k}_{[i} A_{j]}{}^{l]pq}  B_{pq}
-\ft23 A_{[i}{}^{nkl} B_{j]n}
\nonumber\\
&&
-\ft43 \delta_{[i}^{[k} A^{l]}{}_{j]pq}  B^{pq}
+\ft43  A^{[k}{}_{nij} B^{l]n}
-\ft{8}9  \delta^{kl}_{ij} B^{pq} B_{pq}
-\ft{8}9 B^{kl} B_{ij}
+\ft{32}9 B^{n[k} \delta^{l]}_{[i} B_{j]n}
\;,
\nonumber\\[2ex]
{\cal M}_{ijkl} &=&
\ft1{36}  A_{[i}{}^{pqr} \epsilon_{j]pqrmns[k} A_{l]}{}^{mns}
-\ft1{18} \epsilon_{klmnpqr[i} A_{j]}{}^{pqr} B^{mn}
\nonumber\\
&&
+\ft1{9}\epsilon_{ijpqrmn[k} A_{l]}{}^{pqr} B^{mn}
-\ft2{9} \epsilon_{ijklmnpq} B^{mn} B^{pq}
\;.
\eea

Finally, the gravitino and fermion mass matrices are
directly read off from (\ref{eomgrav}) and (\ref{eom:fermions}),
respectively and take the form
\bea
{\cal M}_\psi{}^{ij} &=& \sqrt{2}\left(A^{ij}-2B^{ij}\right)
\;,\nonumber\\[.5ex]
{\cal M}_\chi{}^{ijk,lmn} &=&
\ft1{12} \sqrt{2}\,\left(
  \epsilon^{ijkpqr[lm}\,A^{n]}{}_{pqr}
 +2  \epsilon^{ijklmnpq}  B_{pq} \right)
\;,
\label{Mferm}
\eea
where the first matrix carries the information
about the breaking of supersymmetry and the
latter matrix has to be evaluated after projecting out the fermions
that are eaten by the massive gravitinos.

%
\section{Example: de Sitter geometry and mass spectrum}
\label{sec:example}

We have in the previous section derived the full set of fermionic and bosonic field equations of the gauging in presence of the trombone generator and shown that they transform into each other under supersymmetry. In absence of an action, these equations capture the full dynamics of the theory.
As a simple example and application of the construction, in this section we analyze in more detail the gauging discussed at the end of section~\ref{sttqc}, parametrized by  $\kappa$ and $\Xi^a$\,. In particular, we show that this theory admits a de Sitter solution with constant scalar fields and work out its mass spectrum by linearizing the equations of motion around the vacuum solution.
In the absence of the trombone gauging, i.e.\ for $\kappa=0$,
the theory is characterized by an $E_{6(6)}$ generator $\Xi^a$
and corresponds to the Scherk-Schwarz reduction from five dimensions
first analyzed in~\cite{Cremmer:1979uq,Sezgin:1981ac} and revisited in
the context of four-dimensional gaugings in~\cite{Andrianopoli:2002mf}.

As a first step, we calculate for this theory the value of the tensors
$A_{ij}$, $A_i{}^{jkl}$ and $B^{ij}$ at the origin, i.e.\ for all scalar fields vanishing.
Since at the origin, the group $E_{6(6)}$ is broken down to its
maximally compact subgroup, the values of these tensors will be expressed
in terms of the $USp(8)$ building blocks $(\kappa,  \xi^{ij}, \xi^{ijkl})$,
transforming in the $1$,  $36$, and $42$ of $USp(8)$, respectively,
of which the latter two compose the $E_{6(6)}$ generator $\Xi^a$.
The indices $i, j, \dots = 1, \dots 8,$ here label the fundamental
representation of $USp(8)$. Explicitly, these tensors satisfy the relations
\bea
\xi^{ij}=\xi^{ji}\;,\quad
\xi^{ijkl}=\xi^{[ijkl]}\;,\quad
\xi^{ijkl}\omega_{kl}=0
\;,
\eea
with the $USp(8)$-invariant symplectic matrix $\omega_{ij}$, and the reality properties
\bea
(\xi^{ij})^*= \xi_{ij}=\omega_{ik}\omega_{jl}\xi^{kl}\;,\qquad
\mbox{etc.}
\eea
At the origin, the scalar tensors $A_{ij}$, $A_i{}^{jkl}$ and $B^{ij}$
take the form
\bea
A^{ij} = \ft{1}{\sqrt{2}}
\xi^{ij}
\;,\quad
A_i{}^{jkl} =
-\ft{3}{\sqrt{2}} \omega_{im} \xi^{m[j} \omega^{kl]}
+\omega_{im}\xi^{mjkl}
\;,\quad
B^{ij} =\ft{1}{\sqrt{2}}
\kappa\, \omega^{ij}
\;.
\label{ABk}
\eea
The condition for extremality (\ref{extremum})
coming from the scalar field equations
splits into the equations
\bea
\kappa\,\xi^{ijkl}&=& \sqrt{2}\omega_{mn} \xi^{m[i}\xi^{jkl]n}
\;,\qquad
\xi^{ijkl}\xi_{ijkl}~=~0\;.
\eea
Obviously, even for non-vanishing parameter $\kappa$ these equations leave no other solution than
$\xi^{ijkl}=0$, i.e.\ induce a Scherk-Schwarz gauging with a compact generator
of $E_{6(6)}$\,. On the other hand this shows that choosing $\xi^{ijkl}=0$ suffices to guarantee that the
scalar field equations (\ref{eom:scalars}) are solved by setting all scalar fields to zero.
For the cosmological constant (\ref{coco}), we obtain
\bea
\Lambda
&=&
\ft{3}{8} \xi^{ijkl} \xi_{ijkl}
+ \ft{16}{3} \kappa^2  ~=~ \ft{16}{3} \kappa^2\;,
\eea
i.e.\ the Einstein field equations (\ref{eom:Einstein}) are solved by
a Minkowski space for the standard gaugings and by
a de Sitter geometry for non-vanishing $\kappa$.

The fermionic mass spectrum for this solution is obtained by linearizing the fermionic
field equations (\ref{eomgrav}), (\ref{eom:fermions}) around the de Sitter background
with the mass matrices given by (\ref{Mferm}).
For the eight gravitino masses we obtain
\bea
m_{\rm grav} &:&  \pm\sqrt{m_i^2 +4\kappa^2}\;,\qquad
i=1, \dots, 4
\;,
\eea
where we have denoted by $im_i$ the eigenvalues of the anti-hermitean matrix $\xi_i{}^j$\,.
For non-vanishing $\kappa$ thus all supersymmetries are broken, as
is required by the de Sitter geometry and 8 Goldstinos are found among the
spin-1/2 fermions.
For vanishing $\kappa$ on the other hand, supersymmetry is broken according to
the number of non-vanishing eigenvalues of $\xi_i{}^j$, in accordance with the
results of~\cite{Cremmer:1979uq,Sezgin:1981ac}.

The remaining fermion masses are given by
\bea
m_{\rm ferm} &:&
\pm\sqrt{m_i^2 +4\kappa^2}\;,\quad
\pm\sqrt{(m_i \pm m_j \pm m_k)^2 +4\kappa^2}\;,\;\;
(i<j<k)
\;.
\eea
We find that the effect of the additional trombone gauging is a shift
in all the fermion masses.
Likewise, we may calculate the scalar mass matrix (\ref{Mscalar_sym}), (\ref{Mscalar_alt}),
with (\ref{ABk}) and obtain
\bea
{\cal M}_{ijkl}{}^{mnpq} &=&
\mathbb{P}_{42} \left(
\xi^r{}_s \xi^s{}_r\,\delta^{mnpq}_{ijkl}
-24 \xi^{[m}{}_{[i}\,\xi^n{}_j\,\delta^{pq]}_{kl]}
+32\,\kappa\,\xi^{[m}{}_{[i}\,\delta^{npq]}_{jkl]}
\right)
\mathbb{P}_{42}
\;,
\label{massesC4}
\eea
where $\mathbb{P}_{42}$ refers to the
projector onto the 42 scalars in the decomposition
\bea
{\bf 70} &\longrightarrow& {\bf 1}+{\bf 27}+{\bf 42}
\;,
\eea
of $SU(8)$ under $USp(8)$, i.e.\
all other scalars come with zero mass.
In (\ref{massesC4}), the last term lives entirely in the antisymmetric part
of the mass matrix.
In terms of the eigenvalues of $\xi^m{}_n$, diagonalization of (\ref{massesC4})
leads to the following spectrum
\bea
0&||& 30 \times\nonumber\\
  (m_i \pm m_j)^2 + 4 i \kappa |m_i \pm m_j| &||& i<j \nonumber\\
  (m_i \pm m_j)^2 - 4 i \kappa |m_i \pm m_j|&||& i<j \nonumber\\
(m_1\pm m_2 \pm m_3 \pm m_4)^2 + 4 i \kappa |m_1\pm m_2 \pm m_3 \pm m_4 | &||& \nonumber\\
(m_1\pm m_2 \pm m_3 \pm m_4)^2 - 4 i \kappa |m_1\pm m_2 \pm m_3 \pm m_4 | &||&
\qquad
\;,
\label{spec_scalars}
\eea
for the masses of the scalar fields.
For vanishing $\kappa$, we precisely reproduce the mass spectrum of \cite{Cremmer:1979uq}.
Upon switching on~$\kappa$, all non-vanishing mass-eigenvalues degenerate according to
$m^2 \rightarrow m^2 \pm 4i\kappa m$\,.
The fact that most of the mass eigenvalues come out to be
imaginary is due to the antisymmetric contributions
to the mass matrix.
Finally, the vector mass matrix (\ref{M_vector}) takes the form
\bea
{\cal M}_{ij}{}^{kl} &=&
- \xi^{[k}{}_{[i} \xi^{l]}{}_{j]}
+  \xi^{[k}{}_{n} \delta^{l]}_{[i}\, \xi^n{}_{j]}
+ 4 \kappa \delta^{[k}_{[i} \xi^{l]}{}_{j]}
-\ft{16}{9} \kappa^2 \delta_{ij}^{kl}
-\ft49 \kappa^2 \,\omega_{ij} \omega^{kl}
\;,
\nonumber\\[2ex]
{\cal M}_{ijkl} &=&
\xi_{i[k}\xi_{l]j}
-\ft12\omega_{i[k} \xi^n{}_{l]} \xi_{jn}
+\ft12\omega_{j[k} \xi^n{}_{l]} \xi_{in}
-2\kappa \omega_{i[k} \xi_{l]j} + 2\kappa \omega_{j[k} \xi_{l]i}
\nonumber
\\
&&{}
-\ft{8}9 \kappa^2 \omega_{ij}\omega_{kl}
-\ft{16}9 \kappa^2 \omega_{i[k} \omega_{l]j}
\;.
\label{MMvk}
\eea
In terms of the eigenvalues of the matrix $\xi^m{}_n$, we find the following mass spectrum
\bea
0&||& 28 \times\nonumber\\
  (m_i \pm m_j)^2 + 4 i \kappa |m_i \pm m_j| -\ft{32}{9} \kappa^2 &||& i<j \nonumber\\
  (m_i \pm m_j)^2 - 4 i \kappa |m_i \pm m_j|  -\ft{32}{9} \kappa^2 &||& i<j \nonumber\\
 -\ft{32}{9} \kappa^2  &||& 3 \times \nonumber\\
 -\ft{32}{3} \kappa^2 &||& 1 \times
\;.
\eea
For non-vanishing $\kappa$ thus all 28 vector fields become massive.
The associated massless Goldstone bosons can be identified in the scalar spectrum
(\ref{spec_scalars}) which provides a strong consistency check of the result.
The matrix (\ref{MMvk}) has half-maximal rank in accordance with the fact
that this gauging is purely electric and involves only 28 of the vector fields.

To summarize, we have shown that for the theory discussed at the end of section~\ref{sttqc},
 a de Sitter geometry with constant scalar and
vector fields provides a solution to the full set of field equations. While the
a non-vanishing $\kappa$ in the fermionic sector simply induces a shift in all the fermion
masses, we find that in the bosonic sector, most of the modes have imaginary mass square
eigenvalues. This is due to the fact that the equations of motion do not descend from an action
and may be a sign of an instability of this solution in de Sitter space.
Actually the imaginary shift in the mass squared of the bosonic fluctuations reminds of the Breit-Wigner formula\footnote{We are grateful to Riccardo D'Auria for pointing out this analogy.}
for the propagator of an unstable particle, which has a characteristic imaginary shift in the pole proportional to the particle decay width $\Gamma$:
\begin{eqnarray}
\frac{1}{p^2-m^2+im\Gamma}\,.
\end{eqnarray}
From this point of view our results seem to suggest that the bosonic fluctuations ``die off'' at some characteristic time $\delta t\sim m\Gamma/E$ proportional to the trombone parameter $\kappa$. It would be interesting to understand the implications of this feature for the stability properties of the background. \par
A particular limit of this theory is the case of a `pure trombone gauging', i.e.\ $\xi^m{}_n=0$
with vanishing mass parameters $m_i$\,. It follows from the above formulas, that in this case
all scalar fields remain massless while all vector fields appear with negative mass squares,
again a sign of an instability of the solution. It will be interesting to analyze in more detail, if
this instability is a generic feature of the theories with local scaling invariance or if some classes
of theories among the more complicated gaugings constructed in this paper eventually admit
stable vacuum solutions.

\section{Conclusions}

In this paper, we have derived the most general couplings of four-dimensional supergravity
with a maximal number of supercharges. With a gauge group embedded in the $E_{7(7)} \times \mathbb{R}$ global symmetry group of the Cremmer-Julia theory, the gauge generators are parametrized in terms of an embedding tensor, carrying $56+912$ parameters, subject to a set of bilinear algebraic consistency constraints. After suitable parametrization, we find that the latter reduces to the system (\ref{cs1})--(\ref{cs4}) which allows to construct simple solutions.
The standard gaugings whose gauge group is a subgroup of $E_{7(7)}$ correspond to an embedding tensor in the irreducible ${\bf 912}$ representation. Additional non-vanishing components in the ${\bf 56}$ representation define theories in which local scaling invariance $\mathbb{R}$ (the so-called trombone symmetry) is part of the gauge group.

We have determined the general form of the gauge groups and worked out the full set of modified field equations of these gauged $N=8$ supergravities. As a particular feature of these theories, we have found that a gauging of the trombone generator leads to an additional positive contribution to the effective cosmological constant. Moreover, it turns out that gaugings with local scaling symmetry are generically dyonic, i.e.\ involve simultaneously electric and magnetic vector fields. We have analyzed in detail the simplest example of such a theory which has its higher-dimensional origin as a generalized Scherk-Schwarz reduction from five dimensions. We have shown that this theory admits a de Sitter solution with constant scalar fields and determined its mass spectrum indicating that the solution is not stable.

While in this paper we have analyzed only a single example within the new class of theories, which describes a one-parameter deformation of the known Scherk-Schwarz gaugings from five dimensions,
it would be highly interesting to generalize this analysis to other examples and to perform a systematic study of the possibilities of deformations of the known gaugings. In particular, starting from a theory with supersymmetric AdS vacuum, the additional positive contribution to the effective cosmological constant may lift the space-time geometry to Minkowski or de Sitter upon inclusion of the trombone generator.
In this context it would be important to investigate if the instabilities that we have found in the bosonic spectrum of our example are due to the simple structure of that example or if they persist to more complex situations and represent a generic feature of these theories.

Another interesting aspect for further study is the dyonic structure of the constructed gaugings.
Whereas the appearance of magnetic vector fields is not new and has shown up in previously constructed gaugings in four dimensions~\cite{deWit:2005ub,deWit:2007mt}, in the standard theories there is always a symplectic frame in which all magnetic vector fields drop from the action and the field equations. This frame can be reached in a systematic way by integrating out the two-forms from the action.
In contrast, we have found that for the gaugings constructed in this paper there is in general no symplectic frame in which all gauge fields would be electric. These gaugings are of genuinely dyonic nature.
This does not contradict the no-go results on the gauging of electric/magnetic dualities~\cite{Bunster:2010wv,Deser:2010it}, as the resulting theories do no longer admit an action. It would be highly interesting to study the structure of such dyonic theories in more detail.

\bigskip

It is certainly remarkable that maximal supersymmetry in four dimensions
not only admits the standard gaugings with gauge groups inside $E_{7(7)}$,
described by an embedding tensor in the ${\bf 912}$ representation \cite{deWit:1982ig,deWit:2002vt,deWit:2007mt}, but moreover allows for yet another non-trivial deformation of the
field equations described by 56 additional components of the embedding tensor.
On the other hand, this may be viewed as another sign of the underlying symmetry
structure of extended supergravity theories: upon dimensional reduction to two
dimensions, the global symmetry
group of maximal ungauged supergravity is the affine group $E_{9(9)}$~\cite{Julia:1981wc}
while its gaugings are parametrized by an embedding tensor $\Theta_{\rm 2-dim}$ transforming in the
basic representation of that group~\cite{Samtleben:2007an}.
This infinite-dimensional highest-weight representation
thus captures all deformation parameters of the two-dimensional theory. Decomposition w.r.t.\
the finite-dimensional subgroup $E_{7(7)}\times SL(2)$ gives rise to its lowest level components
\bea
\Theta_{\rm 2-dim} \quad&\longrightarrow&\qquad
\begin{tabular}{r|c}
$+1$&$(1,2)$\\\hline
$+2$&$(56,2)$\\\hline
$+3$&$(133,2)+(1,2)$\\\hline
$+4$&$\;\;\;(912,1)+(56,1)+(56,3)\;\;\;$\\\hline
\dots&\dots
\end{tabular}\quad
\;,
\label{basic}
\eea
from which the higher-dimensional origin of these theories may be inferred.
E.g.\ the theories described by parameters in the
first two rows correspond to torus reductions from four to two dimensions,
in which the KK vector field and the two-dimensional vector fields acquire non-vanishing
flux components along the two-torus, with the corresponding deformation parameters
transforming in the $(1,2)$ and the $(56,2)$, respectively.
Parameters in the third row describe Scherk-Schwarz reduction from four to two dimensions,
including twists with the four-dimensional trombone generator.
The $(912,1)$ in the fourth row corresponds to theories obtained by dimensional reduction
from the standard gaugings in four dimensions, while the $(56,1)$ describes the
dimensional reduction of the theories with local scaling symmetry constructed in this paper.
This shows that after dimensional reduction both the standard gaugings as well as trombone gaugings and combinations of the two are described on equivalent footing by parameters residing within a single irreducible representation of the affine global symmetry group.
In this sense, the new gaugings constructed in this paper may be viewed as obtained by $E_{9(9)}$ rotation from the standard gaugings in four dimensions.
Moreover, the infinite tail of higher level parameters in (\ref{basic}) still advocates the tempting
possibility of discovering yet other maximally supersymmetric couplings in four dimensions
which must however be of genuinely different nature than the present constructions.

\subsection*{Acknowledgments}

The work of H.S.\ is supported in part by the Agence Nationale de la Recherche (ANR).
M.T.\ wishes to thank Riccardo D'Auria for interesting discussions. 
Part of the calculations has been facilitated by use of the computer algebra
system Cadabra~\cite{Peeters:2006kp,Peeters:2007wn}.

\newpage

%
\section*{Appendix}
\begin{appendix}

%
\section{Some algebra}
\label{app:algebra}
%

%
\subsection{Useful $E_7$ relations}
%

We denote by $(t_\alpha)_M{}^N$ the $E_{7(7)}$ generators in the fundamental
representation, i.e.\ the index $\alpha$ runs over $1, \dots, 133$ and $M, N = 1, \dots, 56$\,.
We raise and lower adjoint indices
with the invariant metric $\kappa_{\alpha\beta}\equiv {\rm Tr}\,[t_{\alpha}t_{\beta}]$,
which is a rescaled Cartan-Killing metric.
Fundamental indices are raised and lowered with
the symplectic matrix $\Omega^{MN}$ using north-west south-east conventions:
$X^M=\Omega^{MN}X_N$, etc.\,. We note the following two useful algebraic identities:
\bea
(t^\alpha)_M{}^{K} (t_\alpha)_N{}^{L} &=&
\ft1{24}\delta_M^{K}\delta_N^{L} + \ft1{12}\delta_M^{L}\delta_N^{K}
+(t^{\alpha})_{MN}\,(t_{\alpha})^{KL}-\ft1{24}\Omega_{MN}\,\Omega^{KL}
\;,
\label{relD4}
\eea
and
\bea
(t^\alpha)_{KL} (t_{\alpha})_{MN} &=& \ft1{12} \Omega_{K(M}\Omega_{N)L}+C_{KLMN}
\;,
\eea
with the quartic $E_7$ invariant $C_{KLMN}\equiv(t_\alpha)_{(KL} (t^{\alpha})_{MN)}$\,.

%
\subsection{Breaking $E_{7(7)}$ to $SU(8)$}
%

Upon breaking $E_{7(7)}$ to its maximal compact subgroup $SU(8)$, the fundamental
and the adjoint representation break according to
\bea
{\bf 56} \rightarrow {\bf 28} + {\overline{\bf 28}}\;,\qquad
{\bf 133} \rightarrow {\bf 63} + {\bf 70}
\;,
\eea
respectively. We label the fundamental representation of $SU(8)$ by indices $i, j, \dots = 1, \dots 8$\,.
Then, the $E_{7(7)}$ generators $(t_\alpha)_M{}^N$ break according to
\bea
(t_i{}^j)_{mn}{}^{kl} &=& -\delta^j_{[m} \,\delta^{kl}_{n]i}-\ft18 \delta_i^j \,\delta_{mn}^{kl}
~=~
-(t_i{}^j)^{kl}{}_{mn}
\;,\nonumber\\[1ex]
(t_{ijkl})_{mnpq} &=& \ft1{24}\,\epsilon_{ijklmnpq}
\;,\qquad
(t_{ijkl})^{mnpq} ~=~ \delta_{ijkl}^{mnpq}
\;,
\eea
and the rescaled Cartan-Killing metric
$\kappa_{\alpha\beta}\equiv {\rm Tr}\,[ t_\alpha t_\beta ]$ breaks into
\bea
\kappa_m{}^n,\,_p{}^q &=& 3\left(\delta_m^q\delta_p^n -\ft18 \delta_m^n\delta_p^q\right)
\;,\qquad
\kappa_{ijkl,mnpq} ~=~ \ft1{12} \epsilon_{ijklmnpq}
\;.
\eea

%
\subsection{Breaking $E_{7(7)}$ to $E_{6(6)}\times SO(1,1)$}
%

Upon breaking $E_{7(7)}$ to its maximal subgroup $E_6\times SO(1,1)$,
its lowest dimensional representations decompose according to
\bea
{\bf 56} &\rightarrow&
{\bf 1}^{+3}+{\bf 27}^{+1}+\bar{\bf 27}_{-1}+{\bf 1}^{-3}
\;,
\nonumber\\
{\bf 133} &\rightarrow&
{\bf 1}^{0}+{\bf 78}^{0}+{\bf 27}^{-2}+\bar{\bf 27}^{+2}
\;,
\nonumber\\
{\bf 912} &\rightarrow&
{\bf 78}^{+3}+{\bf 78}^{-3}+{\bf 27}^{+1}+\bar{\bf 27}^{-1}
+{\bf 351}^{-1}+\bar{\bf 351}^{+1}
\;,
\nonumber\\
{\bf 1539} &\rightarrow&
{\bf 1}^{0}+{\bf 78}^{0}+{\bf 650}^{0}+{\bf 27}^{-2}+{\bf 27}^{+4}+\bar{\bf 27}^{+2}+\bar{\bf 27}^{-4}
+{\bf 351}^{+2}+\bar{\bf 351}^{-2}
\;,
\nonumber\\
{\bf 8645} &\rightarrow&
2\cdot{\bf 78}^{0}+{\bf 650}^{0}+{\bf 2925}^{0}
+{\bf 27}^{-2}+\bar{\bf 27}^{+2}
+{\bf 351}^{+2}+{\bf 351}^{-4}
+\bar{\bf 351}^{-2}+\bar{\bf 351}^{+4}
\nonumber\\
&&{}
+{\bf 1728}^{-2}+\bar{\bf 1728}^{+2}
\;,
\label{rep_break}
\eea
with the superscript indicating the $SO(1,1)$ charge.
We use the explicit notation
\bea
{\bf 56}&:& X_M ~\rightarrow~ (X_\bullet,\; X_m,\; X^m,\; X^\bullet)
\;,\nonumber\\
{\bf 133}&:&X_\alpha ~\rightarrow~ (X_{\rm o},\; X_a,\; X_m,\; X^m)
\;,
\eea
with indices $m = 1, \dots, 27$ and $a = 1, \dots, 78$ labeling the
fundamental and the adjoint representation of $E_{6(6)}$, respectively.
The symplectic matrix $\Omega_{MN}$ and the
rescaled Cartan-Killing metric $\kappa_{\alpha\beta}$
break according to
\bea
\Omega_{MN} &\rightarrow&
\left(\Omega_\bullet{}^\bullet=1,\;
\Omega_m{}^n=\delta_m^n,\;
\Omega^\bullet{}_\bullet=-1,\;
\Omega^m{}_n=-\delta^m_n
\right)
\;,
\eea
and
\bea
\kappa_{\alpha\beta} &\rightarrow&
\left(
\kappa_{\rm oo}=72\;,\qquad\kappa_{ab}=2\eta_{ab}\;,\qquad\kappa_m{}^n=12\delta_m^n
\right)
\;.
\label{CK}
\eea
The $E_{7(7)}$ generators $(t_\alpha)_M{}^N$ decompose as
\bea
&&
(t_{\rm o})_\bullet{}^\bullet ~=~3
\;,\qquad
(t_{\rm o})_m{}^n ~=~\delta_m^n
\;,\qquad
(t_{\rm o})^m{}_n ~=~-\delta^m_n
\;,\qquad
(t_{\rm o})^\bullet{}_\bullet ~=~-3
\;,
\nonumber\\
&&
(t_a)_m{}^n ~=~ - (t_a)^n{}_m
\;,
\nonumber\\
&&
(t_m)_\bullet{}^n ~=~ -(t_m)^n{}_\bullet ~=~ \delta_m^n
\;,\qquad
(t^m)^\bullet{}_n ~=~ -(t^m)_n{}^\bullet ~=~ -\delta^m_n
\;,
\nonumber\\
&&
(t_{m})_{nk}~=~ d_{mnk}
\;,\qquad
(t^{m})^{nk}~=~ 10\,d^{mnk}
\;.
\eea
and the decomposition of the structure constants $f_{\alpha\beta}{}^\gamma$
can be read off from the algebra
\bea
&&{}[t_{\rm o}, t_m]=2t_m\;,\qquad
[t_{\rm o},t^m]=-2t^m\;,
\nonumber\\
&&{}[t_a,t_m]=- t_{am}{}^n t_n \;,\qquad
[t_a,t^m]= t_{an}{}^m t^n\;,
\nonumber\\
&&{}[t_m, t^n] = \frac13\,\delta_m^n\,t_{\rm o} -6 (t^a){}_m{}^n\,t_a
\;.
\eea
Here, $d_{mnk}$ denotes the totally symmetric tensor of $E_{6(6)}$,\,
and $ t_{am}{}^n$ denotes the $E_{6(6)}$ generators in the adjoint representation.
Adjoint indices are raised and lowered with the rescaled Cartan-Killing metric
$\eta_{ab}\equiv {\rm Tr}\,[ t_a t_b ]$\,.

%
\subsection{Useful $E_6$ relations}
%

We denote by $d_{mnk}$ and $d^{mnk}$
the totally symmetric tensors of $E_{6(6)}$ in the fundamental ${\bf 27}$ and $\overline{\bf 27}$,
respectively. We choose a relative normalization such that
\bea
d^{mnp}d_{mnq}&=&\delta^p_q\;.
\eea
In the following, we give a list of useful algebraic relations
that be be shown by various contractions and/or by
using an explicit realization of the $E_{6(6)}$ generators:
\bea
d_{mrs}\,d^{spt} \, d_{tnu} \,d^{urq} &=& \ft1{10}\,\delta_{(mn)}{}^{\!\!(pq)}  -
\ft25\, d_{mnr}\, d^{pqr}\,,  \\[1ex]
d_{mps}\,d^{sqt} \, d_{tru} \,d^{upv}\,d_{vqw}\,d^{wrn}  &=& - \ft3{10}
\,\delta_{m}{}^{n}\,,
\\[1ex]
(t^a)_m{}^{k} (t_a)_n{}^{l} &=&
\ft1{18}\,\delta_m^{k}\delta_n^{l} +\ft1{6}\,\delta_m^{l}\delta_n^{k}
-\ft53\, d_{mnp}\,d^{klp}
\;,
\label{relD5}
\\[1ex]
(t_a)_r{}^p (t_b)_s{}^q \,d^{mrs} d_{npq} &=&
-\ft1{30}\,\eta_{ab}\delta^m_n+\ft25 (t_{(a}t_{b)})_n{}^m
\;,
\\[1ex]
(t_a t_b)_r{}^q \,d^{mrs} d_{nqs} &=&
\ft1{30}\,\eta_{ab}\delta^m_n-\ft15 (t_{a}t_b)_n{}^m+\ft3{10}  (t_{b}t_a)_n{}^m
\;,
\\[1ex]
d_{pqr}d^{p(kl}d^{m)qs} &=& \ft1{30} d^{klm} \delta_r^s + \ft1{10} d^{s(kl} \delta^{m)}_r
\;.
\eea
These play a key role in reducing the the $E_{7(7)}$ system of constraints (\ref{Q1})--(\ref{Q3})
for the embedding tensor
to the system (\ref{cs1})--(\ref{cs4}) for its $E_{6(6)}$ components.

\section{Solution of the quadratic constraints}
\label{app:solve}

Our strategy for solving the quadratic constraints
for the embedding tensor follows the analysis of~\cite{LeDiffon:2008sh}
for the pure trombone gaugings. We make use of the fact that under
breaking to $E_{6(6)}$ the tensor $\vartheta_M$ contains a singlet
which (if invertible) allows to explicitly solve all the quadratic equations.
Decomposing under $E_{6(6)}$ according to (\ref{rep_break}),
we label the components $\vartheta_M$ and $\Theta_M{}^\alpha$ of the embedding tensor
according to
\bea
\vartheta_M&\rightarrow~&
(\vartheta_\bullet,\;
\vartheta_m,\;
\vartheta^m,\;
\vartheta^\bullet)
\;.
\label{thbreak}
\eea
and
\bea
\Theta_M{}^\alpha &=&
{\scriptsize
\left(
\begin{array}{cccc}
0& \xi_+^{a}& \xi_n & 0\\
-\ft13\xi_m &
\ft32t^a{}_{m}{}^n\xi_n +3t^a{}_{p}{}^qd_{rqm}\xi^{pr} &
\ft12d_{mnp}\xi^p+\xi_{mn}&
-t_a{}_{m}{}^n \xi_+^{a}\\
-\ft13\xi^m &
\ft32t^a{}_{n}{}^m\xi^n -30t^a{}_{p}{}^qd^{rpm}\xi_{qr} &
-t_a{}_{n}{}^m \xi_-^{a}&
-5d^{mnp}\xi_p+\xi^{mn}
\\
0&\xi^{a}_-&0&\xi^n
\end{array}
\right)}
\;,
\nonumber\\
\label{ThetaBreak}
\eea
respectively, in terms of various $E_{6(6)}$ tensors.
The relative coefficients among the various terms within $\Theta_M{}^\alpha$ are
determined by the fact that $\Theta_M{}^\alpha$ is constrained to live in the ${\bf 912}$
representation of $E_{7(7)}$, i.e.\ satisfies the relations (\ref{linear}).

\subsection{Determining the components of the embedding tensor}

To solve the quadratic constraints in a systematic way, we start from
the equation with the highest $SO(1,1)$ grading.
From (\ref{rep_break}) is follows that this is a ${\bf 27}^{+4}$ representation
inside ${\bf 1539}$, which thus corresponds to evaluating equations (\ref{Q2}) for
$MN={}_m{}_\bullet$. Explicitly, this leads to
\bea
0&\equiv& \xi_m\vartheta_\bullet +t_{am}{}^n \xi_+^a\vartheta_n \;.
\label{eqQQ1}
\eea
Without loss of generality we may assume $\vartheta_\bullet$ to be non-vanishing
(which can always be achieved by change of basis in case $\vartheta_M$ is not
identically zero)
and from (\ref{eqQQ1}) express $\xi_m$
(one of the ${\bf 27}^{+1}$ components of the embedding tensor)
in terms of the unconstrained parameters
$(\vartheta_\bullet, \vartheta_m, \xi_+^a)$
transforming in the ${\bf 1}^{+3}+{\bf 27}^{+1}+{\bf 78}^{+3}$\,.
For convenience, we parametrize the latter as
\bea
\vartheta_\bullet \equiv \kappa\;,\qquad
 \vartheta_m \equiv \kappa \lambda_m\;,
 \qquad \xi^a_+ \equiv \Xi^a
 \;,
 \label{solL1}
\eea
and solve equation (\ref{eqQQ1}) as
\bea
\xi_m &=& - \Xi^a t_{am}{}^n \lambda_n
~\equiv~ - \delta_{\Xi} \lambda_m
\;.
\label{solQQ1}
\eea
By similar computations, the remaining parts of the embedding tensor
can be determined from other components of the constraint equations.
Evaluating equation (\ref{Q3}) for $\alpha\beta={}^m{}^n$
(the $\bar{\bf 351}^{+4}$ equation) yields
 \begin{eqnarray}
0&\equiv&
\vartheta_\bullet\,\xi^{mn}-3\,\Xi^a\,t_{ak}{}^{[m}\,\xi^{n]k}-{10}\,\xi_+^a\,t_{ap}{}^{[m}\,d^{n]pq}\,(\vartheta_q+\ft32\xi_q)\,,
\end{eqnarray}
which upon plugging in (\ref{solQQ1}) reduces to
\bea
{\cal O}_{2/3}\cdot\xi^{mn}
&\equiv&
{10}\,\Xi^a\,t_{ak}{}^{[m}\,d^{n]kl}\: {\cal O}_{2/3}\cdot \lambda_l\,,
\label{eqQQ2}
\eea
where we have defined the operator
\bea
{\cal O}_{2/3} &\equiv& \delta_{\Xi} -\ft23 \kappa
\;.
\eea
As the same operator appears on both sides of equation (\ref{eqQQ2}),
the general solution for the component $\xi^{mn}$ can be given in polynomial form
as
\bea
\xi^{mn} &=& {10}\, \Xi^a t_{ak}{}^{[m}\,d^{n]kl}\,\lambda_l
+\zeta^{mn}
\;,
\label{solQQ2}
\eea
where $\zeta^{mn}$ denotes a (real) zero mode of the operator ${\cal O}_{2/3}$.
It corresponds to an eigenvector of the action of the $\mathfrak{e}_{6(6)}$ generator
defined by $\Xi^a$ with the particular real eigenvalue $\frac23\kappa$\,.
In particular, such zero-modes only exist for non-compact choice of $\Xi^a$ and
some very particular values of $\kappa$\,.

Going down the grading, the next constraint equations live in the
$\bar{\bf 27}^{+2}$, of which there are four different ones. The
relevant ones are obtained from (\ref{Q1}) for $\alpha={}^m$ and
from (\ref{Q2}) for $MN={}^m{}_\bullet$, respectively,
leading to
\bea
(\xi^m+\ft83\vartheta^m)\,\vartheta_\bullet
+\Xi^a\,t_{an}{}^m\,\vartheta^n &=& \xi^{mn}\vartheta_n+
5\,d^{mpq}\vartheta_p\, (\xi_q+\ft83 \vartheta_q) \;,
\label{27a}
\nonumber\\[1ex]
\xi^m\vartheta_\bullet-\Xi^a\,t_{an}{}^m\,\vartheta^n &=&
\xi^{mn}\vartheta_n- 15\,d^{mpq}\vartheta_p\,\xi_q \;.
\label{27b}
\eea
Using the explicit form of (\ref{solQQ1}), (\ref{solQQ2}), these two
equations determine the components $\xi^m$ and $\vartheta^m$
in terms of the free parameters according to
\bea
\vartheta^m &=& 5\kappa\,d^{mkl}\,\lambda_k \lambda_l + \kappa \zeta^m
\;,
\nonumber\\[1ex]
\xi^m &=& 5\,
\Xi^a t_{an}{}^{m}\,d^{nkl}\,
\lambda_k \lambda_l
+ \zeta^{mn} \lambda_n - \ft43 \kappa \zeta^m
\;,
\label{solQQ3}
\eea
up to the constant vector $\zeta^m$ which is a zero mode of the operator
${\cal O}_{4/3} \equiv \delta_{\Xi}-\ft43 \kappa$\,.
Again, such zero modes exist only for very particular values of $\kappa$\,.
Next, we evaluate the constraint (\ref{Q2}) for $MN={}_m{}_n$
(the equation transforming in the ${\bf 351}^{+2}$) to obtain
\bea
0&\equiv&
\ft12\xi_{[m}\vartheta_{n]}-5\,d_{mkp}d^{prq}d_{qln}\,\xi^{kl}\vartheta_r
+t_{a[m}{}^kd_{n]pk}\,\Xi^a \vartheta^p-\xi_{mn}\vartheta_\bullet
\;,
\eea
which uniquely determines the ${\bf 351}^{-1}$ component $\xi_{mn}$
of the embedding tensor.
Explicitly, after some computation and using the relations obtained above, we find
\bea
\xi_{mn} &=& 2\,\Xi^a t_{a[m}{}^k\,\lambda_{n]}  \lambda_k
-5\, \Xi^a t_{a[m}{}^p d_{n]pq}d^{qkl}\,\lambda_k\lambda_l
\nonumber\\
&&{}
-5\,d_{mkp}d^{prq}d_{qln}\,\zeta^{kl}\lambda_r
+t_{a[m}{}^kd_{n]pk}\,\Xi^a\zeta^p
\;.
\label{solQQ4}
\eea
The singlet equation ${\bf 1}^0$ from evaluating
(\ref{Q1}) for $\alpha={\rm o}$  yields
\bea
\ft43 \vartheta_\bullet \vartheta^\bullet &=&
\ft13\,(\xi^m\vartheta_m-\xi_m\vartheta^m) -\ft49\, \vartheta^m\vartheta_m
\;,
\eea
which allows to express the singlet ${\bf 1}^{-3}$ component $\vartheta^\bullet$ in terms of the
other fields
\bea
 \vartheta^\bullet &=&
 -\ft{5}{3}\,\kappa \, d^{k l m} \, \lambda_k \lambda_l \lambda_m
  - \kappa  \zeta^m \lambda_m
 \;.
 \label{solL2}
\eea
Evaluating (\ref{Q1}) for $\alpha=a$ finally yields
\bea
0&=&
\ft32 t^a{}_{m}{}^n (\xi^m\vartheta_n-\xi_n\vartheta^m)
-3 t^a{}_{p}{}^q
(d_{rqn}\xi^{pr}\vartheta^n+10 d^{rpn}\xi_{qr}\vartheta_n)
\nonumber\\[.5ex]
&&{}
+\xi_-^a\vartheta_\bullet-\Xi^a\vartheta^\bullet
+16t^a{}_{m}{}^n\vartheta^m\vartheta_n
\;.
\eea
which yields
the ${\bf 78}^{-3}$ component $\xi_-^a$
\bea
\xi_-^a &=& 30\,\xi^b_+ (t_b t^a)_n{}^m\,d^{nkl} \,\lambda_{m}\lambda_{k}\lambda_{l}
-\ft53 \,\Xi^a\,d^{klm}\,\lambda_{m}\lambda_{k}\lambda_{l}
\nonumber\\
&&{}
+ \Xi^a \lambda_k \zeta^k
- 6 \Xi^b (t^a t_b)_k{}^l \lambda_l \zeta^k
-8 \kappa\,  t^a{}_k{}^l \lambda_l \zeta^k
\nonumber\\
&&{}
-6 t^a{}_m{}^l \lambda_k\lambda_l \zeta^{km}
+15 t^a{}_n{}^q d^{klp}d_{mpq} \lambda_k \lambda_l \zeta^{mn}
\;.
\eea
We have thus determined all the components of the embedding tensor (\ref{thbreak}), (\ref{ThetaBreak}) in terms of the parameters $\kappa$, $\lambda_m$, $\Xi^a$, $\zeta^m$, $\zeta^{mn}$, of which the latter two are particular eigenvectors under the $E_{6(6)}$ action of $\Xi$\,.
In the following, we need to check that this solution indeed satisfies all the
constraint equations~(\ref{Q1})--(\ref{Q3}).
In particular, the remaining constraint equations may impose further restriction on these tensors.
E.g.\ evaluating equation~\ref{Q1}) for $\alpha={}_m$ implies that
\bea
d_{mkl} \zeta^k \zeta^l &=& 0
\;,
\label{zetzet}
\eea
i.e.\ $\zeta^k$ is not only zero mode of ${\cal O}_{4/3}$ but also
satifies a `pure-spinor type' condition of $E_{6(6)}$\,.
In the following, we evaluate all the remaining constraint equations.

\subsection{Evaluating the remaining equations}\label{etre}

We can now check all the remaining equations upon using the solution obtained above.
To this end, let us first calculate the invariant $I_4(\vartheta)$ from (\ref{I4}) quartic in the
trombone parameters $\vartheta_M$ for the explicit solution (\ref{solL1}), (\ref{solQQ3}), (\ref{solL2}).
As a result, we obtain
\begin{eqnarray}
I_4(\vartheta)&=&\frac{2}{3}
\,\kappa^4\,d_{mnp}\,\zeta^m\zeta^n\zeta^p\,,
\end{eqnarray}
i.e.\ the quartic invariant does not depend on the parameters $\lambda_m$\,.
This shows that all $\lambda_m$ can be set to zero by an $E_{7(7)}$ transformation
and therefore do not induce inequivalent gaugings. For simplicity, we will thus in the following
set $\lambda_m=0$\,.
The solution found in the previous section then reduces to
the solution (\ref{solutionemb}) given in the main text. In this section
we will evaluate all remaining constraint equations for this solution.
The calculation is rather tedious and has been performed using mathematica and
cadabra~\cite{Peeters:2006kp,Peeters:2007wn}.
As a result, we find that all remaining constraint equations of the system (\ref{Q1})--(\ref{Q3})
are satisfied provided, the parameters $\zeta^k$, $\zeta^{mn}$
obey the following set of identities
\bea
 \zeta^k\zeta^l d_{mkl} &=& 0
\;,\label{ps1}\\
\zeta^{k} \zeta^{mn} d_{kml} &=& 0\;,
\label{ps2}\\
\zeta^{[k} \zeta^{mn]} &=& 0 \;,
\label{ps3}\\
\left(t_a \cdot (\Xi+\ft43\kappa I) \cdot (\Xi-\ft23\kappa I)\right){}\!_n{}^m \,\zeta^n &=&
-\ft12  \zeta^{mk} \zeta^{ln} d_{klp} (t_a)_n{}^p
\;,
\label{remaining_constraints}
\eea
with the matrix $\Xi$ given by $\Xi_m{}^n \equiv \Xi^a (t_a)_m{}^n$ and `$\cdot$'
denoting the matrix product.
The third equation
comes from the constraint (\ref{Q3}) evaluated for $[\alpha\beta]=[ab]$;
the fourth equation comes from
the same constraint evaluated for $[\alpha\beta]=\, ^{am}$.
We have explicitly verified that {\em all other constraints}
 are satisfied as a consequence of the ansatz
and the relations (\ref{remaining_constraints}).
In particular, the constraint obtained from (\ref{Q3}) evaluated for $[\alpha\beta]=\, ^{a}{}_{m}$
follows after some computation from
the fourth equation of (\ref{remaining_constraints}) and the other constraints.

The fourth equation of (\ref{remaining_constraints}) is linear in $\zeta^n$.
In particular, contracting this equation with $(t^a)_m{}^q$ implies
\bea
(32\,\kappa^2- \Xi^a\Xi_a)\,\zeta^m &=& \ft{15}{2}\, \zeta^{kl}\zeta^{rs} d_{krp}d_{lsq} d^{pqm}
\;,
\label{contract4}
\eea
i.e.\ for $\Xi^a\Xi_a\not=32\,\kappa^2$, we can express $\zeta^m$ as a bilinear in $\zeta^{kl}$\,.
Note that this is consistent, as the r.h.s.\ of (\ref{contract4}) is indeed eigenvector of $\Xi$ associated to
an eigenvalue which is twice the one of $\zeta^{kl}$\,.
Still, plugging this expression for $\xi^m$ into the four equations (\ref{remaining_constraints})
will lead to a set of nontrivial constraints polynomial in $\zeta^{mn}$\,.
\par
As illustrated in section \ref{awoe}, the ``$ {\rm E}_{6(6)}$-pure
spinor'' constraint (\ref{ps1}) on $\zeta^m$ singles out an ${\rm
O}(1,1)\times{\rm SO}(5,5)$ subgroup of ${\rm E}_{6(6)} $ in which
${\rm SO}(5,5)$ is part of the little group of its solution. In
particular $\zeta^m$ coincides with the singlet ${\bf 1}^{-4}$ in
the branching (\ref{o11o55branch}) of the $\overline{{\bf
27}}$ relative to this ${\rm O}(1,1)\times{\rm SO}(5,5)$, and $\Xi$ should have the form in Eq. (\ref{xisemisimple}).
Let us first consider the simple case in which $\Xi_0$ is a
semisimple element of $\mathfrak{so}(5,5)$ and thus can be
considered as an element of its Cartan subalgebra. Let us also
consider the case in which the two sides of Eq. (\ref{cs4}) are
separately zero. The constraint (\ref{remaining_constraints}) is
then satisfied if $\Xi_0$ commutes with an $\mathfrak{so}(4,4)$
subalgebra of $\mathfrak{so}(5,5)$ and if its norm is ${\rm
Tr}(\Xi_0\cdot \Xi_0)=24\,\kappa^2$. The last requirement is easily
understood by observing that $\Xi_0$ and $\Xi_1$ are mutually
orthogonal and that ${\rm Tr}(\Xi_1\cdot \Xi_1)=8\,\kappa^2$, since
(\ref{remaining_constraints}) implies that ${\rm Tr}(\Xi\cdot
\Xi)=32\,\kappa^2$. This fixes the normalization of $\Xi_0$. We can
branch the relevant ${\rm E}_{6(6)}$ representations with respect to
its ${\rm SO}(1,1)^2\times {\rm SO}(4,4)$ subgroup, where the ${\rm
SO}(1,1)^2$ factor is generated by $\Xi_1+\Xi_0$. The
$\overline{{\bf 27}}$ then branches as follows:
\begin{eqnarray}
\overline{{\bf 27}}&\rightarrow & {\bf 1}^{(\frac{4}{3},0)}+{\bf 1}^{(-\frac{2}{3},2)}+{\bf 1}^{(-\frac{2}{3},-2)}+{\bf 8}_v^{(-\frac{2}{3},0)}+{\bf 8}_s^{(\frac{1}{3},1)}+{\bf 8}_c^{(\frac{1}{3},-1)}\,,\nonumber\\
{{\bf 78}}&\rightarrow & ({\bf 28}+{\bf 1}+{\bf 1})^{(0,0)}+{\bf 8}_s^{(1,-1)}+{\bf 8}_c^{(1,1)}+{\bf 8}_s^{(-1,1)}+{\bf 8}_c^{(-1,-1)}+
{\bf 8}_v^{(0,2)}+{\bf 8}_v^{(0,-2)}\,,\nonumber
\end{eqnarray}
where the gradings are the eigenvalues of $\Xi_1/\kappa,\,\Xi_0/\kappa$. The ``pure spinor'' $\zeta^m$ corresponds to  the ${\bf 1}^{(\frac{4}{3},0)}$ representation. Consider now the vector $\zeta\cdot t^a \equiv(\zeta^m\,t^a{}_m{}^n)$. The constraint (\ref{remaining_constraints}) reads:
\begin{eqnarray}
\left(\delta_\Xi+\frac{2}{3}\,\kappa\right)\left(\delta_\Xi-\frac{4}{3}\,\kappa\right)\zeta\cdot t^a =0\,.
\end{eqnarray}
Let us analyze the relevant cases:
\begin{itemize}\item{If $t^a\in \mathfrak{so}(5,5)+\mathfrak{so}(1,1)$, $\zeta\cdot t^a$ is still in the ${\bf 1}^{(\frac{4}{3},0)}$ and is thus annihilated by $\delta_\Xi-\frac{4}{3}\,\kappa$;}
\item{If $t^a\in {\bf 8}_s^{(1,-1)}+{\bf 8}_c^{(1,1)}$, $\zeta\cdot t^a={\bf 0}$ and the constraint is satisfied;}
  \item{If $t^a\in {\bf 8}_s^{(-1,1)}+{\bf 8}_c^{(-1,-1)}$, $\zeta\cdot t^a\in {\bf 8}_s^{(\frac{1}{3},1)}+{\bf 8}_c^{(\frac{1}{3},-1)}$. The component in ${\bf 8}_s^{(\frac{1}{3},1)}$ is a zero-mode of  $\delta_\Xi-\frac{4}{3}\,\kappa$, while the second component is a zero mode of $\delta_\Xi+\frac{2}{3}\,\kappa$ and the constraint is still satisfied;}
\end{itemize}
Consider now the case in which $\Xi_0$ has a nilpotent component $\Xi_{\rm n}$ in ${\bf 16}_s^{-3}$, so that $\Xi$ has a semisimple and a nilpotent component $\Xi=\Xi_{\rm ss}+\Xi_{\rm n}$. The constraint (\ref{remaining_constraints}) is still satisfied provided the following condition holds: $[\Xi_{\rm ss},\Xi_{\rm n}]=2\,\kappa\,\Xi_{\rm n}$.

\section{Gauge group generators in $E_6$ components}

Here, we give the gauge group generators $(X_M)_N{}^K$
for the general solution (\ref{solutionemb}) of the quadratic constraints as obtained from
(\ref{defX}), (\ref{thbreak}), (\ref{ThetaBreak}), and (\ref{solutionemb}).
They are parametrized by $\kappa$, $\Xi^a$, $\zeta^m$, and $\zeta^{mn}$,
which are subject to the identities (\ref{remaining_constraints}).
\bea
X_\bullet &=&
{\scriptsize
\left(
\begin{array}{cccc}
0&0&0&0\\
0&\Xi_m{}^n-\ft23 \kappa \delta_m^n&0&0
\\
0&0& -\Xi_n{}^m-\ft43 \kappa \delta_n^m&0
\\
0&0&0& -2\kappa
\end{array}
\right)}
\;,
\nonumber\\[1ex]
X_k &=&
{\scriptsize
\left(
\begin{array}{cccc}
0&-\Xi_k{}^n+\ft23 \kappa \delta_k^n&0&0\\
\Xi_{[k}{}^p \zeta^q d_{m]pq}
&\ft12
\zeta^{np}d_{kpm}+5\zeta^{pq}d_{kpr}d_{qsm}d^{rsn}
&\ft23\kappa d_{kmn}-\Xi_{k}{}^pd_{pmn}
&0
\\
0&10\Xi_{[k}{}^p\zeta^q d_{r]pq}d^{rmn}&
\hspace*{-5mm}
5\zeta^{pq}d_{npr}d_{kqs}d^{rsm}-\ft12\zeta^{mp}d_{kpn}
&
\Xi_k{}^m-\ft23 \kappa \delta_k^m
\\
0&0&
-\Xi_{[k}{}^p \zeta^q d_{n]pq}
& 0
\end{array}
\right)}
\;,
\nonumber\\[1ex]
X^k &=&
{\scriptsize
\left(
\begin{array}{cccc}
0&\zeta^{kn}&0&0
\\
0& X^k{}_m{}^n & \zeta^{kp} d_{pmn} &0
\\
0&0& X^{k\,m}{}_n & \zeta^{mk}
\\
0&0&0&-2\kappa\,\zeta^m
\end{array}
\right)}
\;,
\nonumber\\[1ex]
 X^\bullet &=& {\scriptsize \left(
\begin{array}{cccc}
0&-2\kappa\zeta^n&0&0\\
0&0&-2\kappa \zeta^p d_{pmn}&0
\\
0&0&0&2\kappa\zeta^m
\\
0&0&0&0
\end{array}
\right)} \;, \label{explicitgenerators} \eea
where
\begin{eqnarray}
X^k{}_m{}^n&=& 2\Xi_m{}^{[k}\zeta^{n]} -\ft43\kappa
\zeta^k\delta^n_m -\ft23 \kappa  \zeta^n\delta^k_m+\ft{20}3 \kappa
\zeta^p d_{pqm}d^{qkn} -10 \Xi_p{}^k\zeta^q d_{mqr}d^{rpn}\;,
\nonumber\\
X^{k\,m}{}_n&=&2\Xi_n{}^{[m}\zeta^{k]}+\ft43\kappa
\delta_n^{[k}\zeta^{m]} -\ft{20}3 \kappa  \zeta^p d_{pqn}d^{qkm} +10
\Xi_p{}^k\zeta^q d_{nqr}d^{rpm} \;. \nonumber
\end{eqnarray}
Via (\ref{XXX}) these generators also encode the structure constants of the
gauge algebra.

\mathon\section{$T$-identities}
\mathoff
\label{app:Tids}

Upon dressing the quadratic constraints (\ref{Q1})--(\ref{Q3}) with the scalar vielbein
and using the definitions (\ref{TX}), (\ref{TAB}), one obtains a large number of $SU(8)$
identities bilinear in the tensors $A$ and $B$.
Here we collect those identities that are important for the calculations in the main text.
Working out all quadratic constraints
that transform in the ${\bf 63}$ of $SU(8)$, we find the following relations
\bea
{\bf 1539}: \;\; 0  &=&
6 A^{ik}B_{jk}-6 B^{ik}A_{jk}+A_j{}^{imn}B_{mn}-A^i{}_{jmn}B^{mn}
\;,
\label{constraints63}
\\[1ex]
{\bf 133}:\;\; 0 &=&
2 A^{ik}B_{jk}+2 B^{ik}A_{jk}-A_j{}^{imn}B_{mn}-A^i{}_{jmn}B^{mn}
+\ft{32}{3} B^{ik}B_{jk}
~-~ \ft18 \delta^i_j \,{\rm trace}
\;,
\nonumber\\[1ex]
{\bf 133}: \;\; 0  &=&
12 A^{ik}A_{jk}-A^i{}_{mnk} A_j{}^{mnk} +3 A^k{}_{mnj}A_k{}^{mni}
\nonumber\\
&&{}
+ 12
\left(2 A^{ik}B_{jk}+2 B^{ik}A_{jk}-A_j{}^{imn}B_{mn}-A^i{}_{jmn}B^{mn}\right)
~-~ \ft18 \delta^i_j \,{\rm trace}
\;,
\nonumber\\[1ex]
{\bf 8645}: \;\; 0  &=&
-240 A^{ik}A_{jk}+11A^i{}_{mnk} A_j{}^{mnk} +21 A^k{}_{mnj}A_k{}^{mni}
\nonumber\\
&&{}
-12
\left(10 A^{ik}B_{jk}+10 B^{ik}A_{jk}+A_j{}^{imn}B_{mn}+A^i{}_{jmn}B^{mn}\right)
~-~ \ft18 \delta^i_j \,{\rm trace}
\;,\nonumber
\eea
descending from the various irreducible $E_{7(7)}$ contributions of (\ref{Q1})--(\ref{Q3})
as indicated. The $E_{7(7)}$ origin of these constraints can also be confirmed by calculating
the action of the quadratic Casimir operator upon using the $E_{7(7)}$ transformation properties
\bea
\delta A^{ij} &=&
\ft13 A^{(i}{}_{klm} \Sigma^{j)klm}
\;,
\nonumber\\
\delta A_i{}^{jkl} &=&
2A_{im} \Sigma^{mjkl} + 3\Sigma^{mn[jk} A^{l]}{}_{imn}
+\Sigma^{mnp[j}\delta^k_i A^{l]}{}_{mnp}
\;,
\nonumber\\
\delta B^{ij} &=&
-\Sigma^{ijkl}B_{kl}
\;,
\label{actionE7}
\eea
with $\Sigma^{ijkl}$ satisfying
$\Sigma_{ijkl}=\frac1{24}\epsilon^{ijklmnpq}\Sigma_{mnpq}$\,.

Similarly, we can deduce
we can deduce two constraints in the
${\bf 70}$ of $SU(8)$:
\bea
{\bf 133}: \;\; 0  &=&
A^m{}_{[jkl} B_{n]m}
+4B_{[jk}B_{ln]}
  -\ft1{24} \epsilon_{jkln mqrs}
  \left(A_p{}^{mqr} B^{sp}+4B^{mq}B^{rs} \right)
\nonumber\\[1ex]
{\bf 133}: \;\; 0   &=&
 4 A^m{}_{[jkl} A_{n]m}
  -3  A^m{}_{p[jk}A^p{}_{ln]m}
  +16 A^m{}_{[jkl} B_{n]m}
  \nonumber\\
  &&{}
  -\ft1{24}  \epsilon_{jklnmpqr} \left(
  4A_s{}^{mpq}A^{rs}
  -3A_t{}^{ump} A_u{}^{qrt}
  +16 A_u{}^{mpq} B^{ru}\right)\;,
\label{constraints70}
\eea
and two constraints in the
${\bf 378}$ of $SU(8)$:
\bea
{\bf 1539}: \;\; 0  &=&
4A_{j[k}B_{ln]} + A^m{}_{j[kl} B_{n]m} + A^m{}_{kln} B_{jm}
\nonumber\\
&&{}
-\ft1{9} \epsilon_{rsmpqnkl} A_j{}^{mpq} B^{rs}
+\ft1{18} \epsilon_{mjqrsnkl} A_p{}^{qrs} B^{mp}
\;,
\nonumber\\[1ex]
{\bf 8645}: \;\; 0  &=&
-18 A^m{}_{nkl}A_{jm} - 54 A^m{}_{j[kl} A_{n]m}
+60A_{j[k}B_{ln]} -9 A^m{}_{j[kl} B_{n]m} -9 A^m{}_{kln} B_{jm}
\nonumber\\
&&{}
+  \epsilon_{mpqrsnkl} A_j{}^{urs}
(A_u{}^{mpq}
-\ft1{3} \delta_u^m B^{pq} )
-\ft34 \epsilon_{jnklpmrs} A_u{}^{rst}
(A_t{}^{mup} -\ft2{9} \delta^m_t B^{up})
\;.
\nonumber\\
\label{constraints378}
\eea
If the components $\Theta_M{}^\alpha$, $\vartheta_M$ are chosen such
as to satisfy the quadratic constraints (\ref{Q1})--(\ref{Q3}), the relations
(\ref{constraints63})--(\ref{constraints378}) among the scalar tensors $A$, $B$,
follow as an immediate consequence.
For $B^{ij}=0$, all these identities consistently reduce to the quadratic identities
given in~\cite{deWit:1982ig,deWit:2007mt}.

\end{appendix}



\providecommand{\href}[2]{#2}\begingroup\raggedright\endgroup

\end{document}